\documentclass[a4paper,12pt]{article}
\pdfoutput=1
\usepackage[utf8]{inputenc}
\usepackage{graphicx}
\usepackage{caption}
\usepackage{subcaption}
\usepackage{amsmath,amssymb}
\usepackage[margin=0.80in]{geometry}
\usepackage{color}
\usepackage[usenames,dvipsnames]{xcolor}

\title{Collective behavior of viscoelastic asperities as a model for static 
       and kinetic friction}
\author{Srivatsan Hulikal, Kaushik Bhattacharya, Nadia Lapusta\\
 Department of Mechanical and Civil Engineering\\
 California Institute of Technology\\
 Pasadena CA 91125}
\date{}

\begin{document}

\maketitle

\begin{abstract}
We propose a statistical model for static and sliding friction
between rough surfaces. Approximating the contact between rough surfaces
by the contact of an ensemble of  one-dimensional viscoelastic 
elements  with a rough rigid surface, we study the 
collective behavior of the elements.  We find that collective response of 
the contacts can lead to macroscopic 
behavior very different from the microscopic behavior.
Specifically, various observed features of friction emerge as collective
phenomena, without postulating them directly at the microscale.
We discuss how parameters in our model can be
related to material and surface properties of the contacting surfaces.
We compare our results to commonly used rate and state phenomenological models,
and propose a new interpretation of the state variable.
\end{abstract}

\maketitle

\section{Introduction}

Friction between surfaces plays an important role in phenomena spanning
many length scales, and in diverse fields including engineering, biology and 
geology \cite{urbakh:1,dieterich:9}. Friction is ubiquitous: it allows us to 
walk and drive, and it plays a key role in the working of many machines and 
technologies. 
At the same time, losses due to friction and wear amount
to a significant fraction of the GNP \cite{jostHP:1,bhushanB:2}. Thus, the
study of friction potentially entails great economic benefits. At small length 
scales, the ratio of surface area to volume being large, surface forces 
play a dominant role. Hence, in the design of small scale technologies like 
MEMS, NEMS and magnetic disk drives, friction has to be given a careful 
consideration \cite{tambeNS:1,mcfadyen:1,bhushanB:1}. Another application of 
the study of friction is tactile sensing, where the goal is to endow 
machines with a sense of touch \cite{scheibertJ:1,wandersmanE:1}.
Various aspects of earthquakes are known to be sensitive to the frictional
properties on faults \cite{scholz:1,marone:1,dieterich:9}.  
For these reasons, in the last few decades, 
there has been a resurgence in interest in friction which, accompanied by the 
development of new experimental techniques and increased computational power, 
has resulted in a 
number of studies of frictional properties of interfaces in 
different materials at different length and time scales. 

The classical picture of friction that emerged from the studies of 
Leonardo Da Vinci, Guillaime Amontons, and Agustus Coulomb among others is: 
(a) friction between surfaces is characterized by two numbers, a static friction 
coefficient $\mu_s$ and a kinetic friction coefficient $\mu_k$. $\mu_s$ is the 
ratio between the shear force required to initiate sliding and the normal force,
and $\mu_k$ is (per unit normal force) the shear force 
necessary to sustain sliding at a constant (nonzero speed) 
velocity, (b) the friction coefficients $\mu_s$ and $\mu_k$ are independent of
the normal force applied and nominal area of the sliding 
surfaces, and (c) $\mu_k$ is independent of the sliding speed \cite{dowson:2}. 
Careful experiments on macroscopic systems have shown, however, that $\mu_s$ 
depends on how long surfaces are held in contact before sliding is induced, 
and $\mu_k$ depends on the sliding speed \cite{rabinowiczE:1,dieterich:2,
marone:1,scholz:2,beeler:1}. 
Further, the frictional 
resistance depends not only on the current sliding velocity, but also on the 
velocity history of the system \cite{dieterich:2,ruina:1,marone:1,beeler:1}. 
The independence of the friction coefficients
with respect to the normal force and the nominal area of contact has been 
observed to be a good approximation, except when the 
normal force varies rapidly \cite{prakash:1,linker:1}.
Section \ref{sec:Background} describes some of the experimental results and
a class of empirical rate and state laws that has been used to model frictional
behavior.

\begin{figure}
\centering
\includegraphics[scale=0.4]{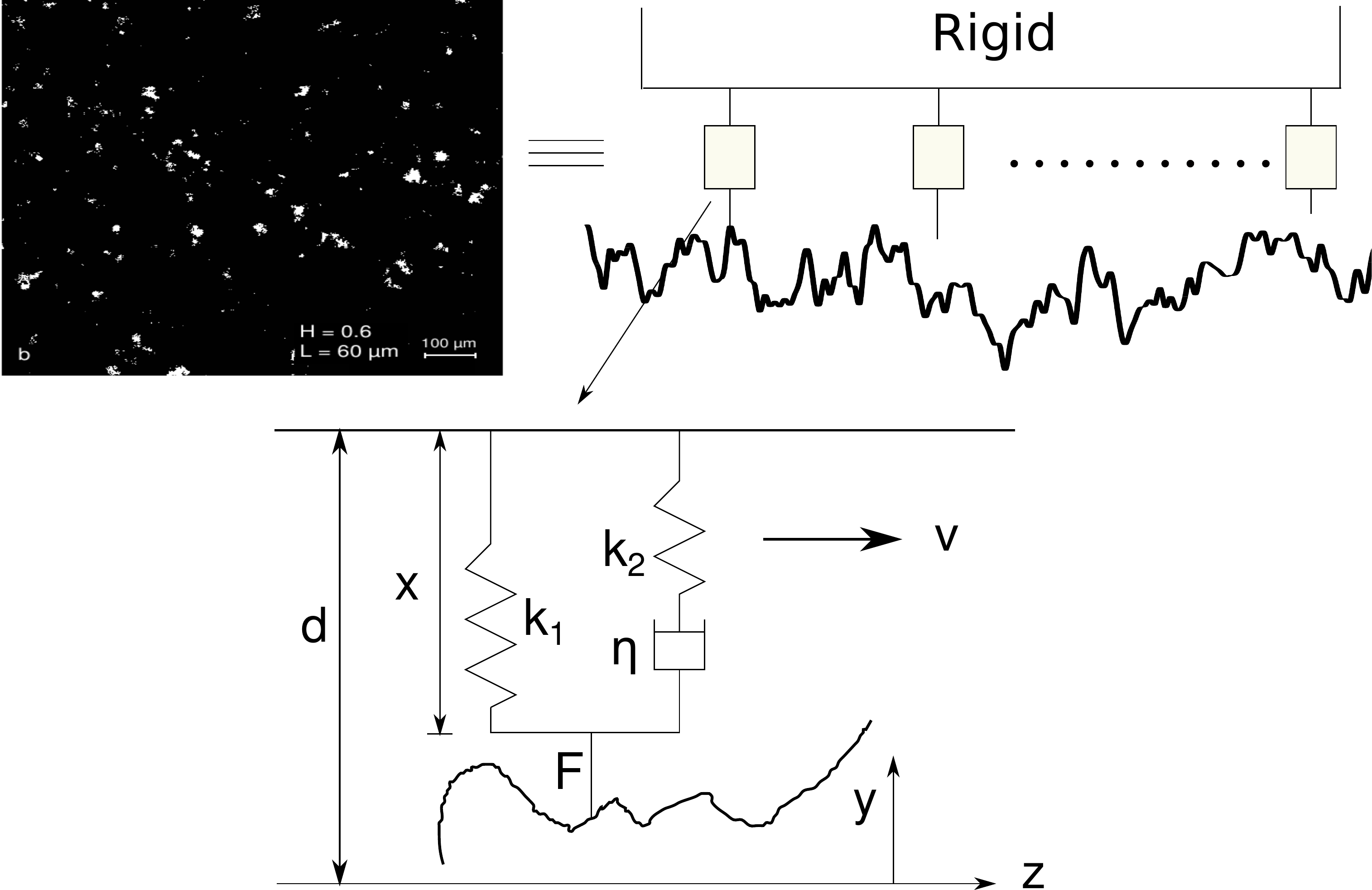}
\caption{(Top left) Microscale image of actual contacts (white spots) between two 
         rough surfaces (adapted with permission from \cite{dieterich:10}). The
         contacts form at the peaks (asperities) of the surfaces.
         (Top right) We model the system as an ensemble of one-dimensional
elements in contact with a rigid rough surface. (Bottom) Each asperity  is 
represented by a viscoelastic spring-dashpot element.}
\label{fig:sls_system}
\end{figure}

Various theories of contact between surfaces have been proposed. 
Most surfaces, even those that appear smooth, are rough at the microscale. 
When two such rough surfaces are pressed against each other, actual contact 
occurs only at a few spots, at the peaks (asperities) of the surfaces (Figure 
\ref{fig:sls_system}).  There is a large body of literature on single asperity
contact, starting from the problem of elastic contact between spheres first 
addressed by Hertz to theories that include plasticity and adhesion 
\cite{bhushanB:3,johnsonKL:1,derjaguinBV:1}.

Two broad classes of models have been proposed to connect the asperity scale to
the experimentally observed features of the macroscopic frictional behavior.  
In one class, the contacts are considered to be plastic, following Bowden and 
Tabor \cite{bowdenFP:1} who suggested that, because of surface roughness,
the actual area of contact is only a small fraction of the nominal area, 
and high local stresses often reach the yield stresses of the materials. 
They estimated the coefficient of friction as the ratio of shear strength of
contacts to the indentation hardness of the material:
$$ F_N = A_r \sigma_c, \quad F_S = A_r \tau_c, 
\quad \mu = F_S/F_N = \tau_c/\sigma_c,$$
where $F_N$ and $F_S$ are the macroscopic normal and shear loads, $A_r$ is the 
real area of contact, $\sigma_c$ is the indentation hardness, $\tau_c$ 
is the asperity shear strength, and $\mu$ is the friction coefficient.
Several subsequent studies have incorporated the time and velocity dependence 
into that framework by representing the shear force at a contact as the 
product of the contact shear strength that depends on the sliding velocity and
area that depends on the age of the contact \cite{brechetY:1,estrin:1,
berthoudP:1,baumberger:2,putelat:1}.
The velocity dependence of the shear strength is attributed to an
Arrhenius type activation mechanism while the time dependence of the area 
results from the creep behavior of the material. 
The proposed formulations have been able to match various friction observations.
In these models, it is assumed that each contact has the same shear and normal
force per unit area and the evolution of the contact population is 
accounted for only by the evolution of the total contact area. 
As the total contact area changes, the normal force per unit area adjusts, 
providing the only interaction between the macroscale and the single asperity.
Hence this class of models is dominated by the behavior of single asperities and
does not include the effects of the statistical properties of the contacting 
rough surfaces.

In the other class of models, the contacts are considered to be elastic.  
Since the shear and normal forces are no longer proportional at the microscale 
for this case, the collective behavior of asperities becomes paramount in 
explaining the proportionality at the macroscale. Archard \cite{archard:1} 
proposed a hierarchical model in which each elastic contact is made
of multiple contacts at a smaller scale. 
This is a precursor to the fractal models of contact
\cite{majumdarA:1,persson:2,ciavarella:1}. 
In the model proposed by Greenwood and Williamson (GW) \cite{greenwood:1}, each
asperity is assumed to be spherical and a single contact is modeled according to
Hertzian theory. This is fitted within a statistical description of the rough
surface. These models are capable of explaining the basic observations of 
proportionality between the shear and normal forces at the macroscale and 
hence the constant static coefficient. They ignore however, the spatial 
features of surface roughness and time-dependent behavior of single asperities.
Thus they are unable to explain any evolution of the friction coefficient.
 
Our study is the first step towards bridging the gap between the two classes of
models and examining the interaction between time-dependent behavior of single
asperities and statistical properties of the rough contacting surfaces.  
When two surfaces are held in static contact or slid against each other, 
the population of contacts evolves: contacts grow, become smaller, and come into
and go out of existence.  Concomitantly, the forces on the asperities also 
evolve, not only because of the time dependence of each contact, but also 
because of the statistical properties of the rough contacting surfaces.
This evolution at the microscale results in the evolution of friction, normal 
force, area of contact etc., at the macroscale.  
To explore the interaction between these two influences, we represent the system
of two surfaces in contact by an ensemble of  one-dimensional viscoelastic 
elements in contact with a rough rigid surface.  
Combining the ideas of the two classes of models, we assume that the local 
shear force between a contacting element and the rough surface
depends both on the sliding velocity and the local
normal force. The state of the system at any instant
is described by a probability distribution of normal forces 
experienced by the elements in the ensemble.
Any macroscopic variable can be determined by integrating the 
appropriate quantity  with respect to this probability distribution. 
The global normal force, which is the sum of the normal forces on
all contacts,  is the first moment with respect to the probability 
distribution.  The macroscopic friction is obtained as an appropriate
moment (depending on the local friction law) of the normal force 
distribution.  
Therefore, the time, velocity, and history dependence of frictional properties 
at the macroscale are a manifestation of the evolution of the probability 
density. Consequently, the behavior at the 
macroscale can be very different from the behavior of single asperities at the 
microscale. For example, even if individual asperities have
velocity-strengthening local friction (higher resistance when sliding faster), 
the macroscale behavior can be velocity weakening (lower resistance when sliding 
faster). 

We describe our multiscale model for friction, discuss its properties, 
and compare its results with experiments in Section 
\ref{sec:Viscoelastic asperity model}. 
We also compare our results to commonly used rate and state phenomenological 
laws, and propose a new interpretation for the state variable that enters the 
laws. In Section \ref{sec:Conclusion}, we summarize our findings and 
suggest ways in which the model can be improved.

\section{Background}
\label{sec:Background}

\subsection{Static friction coefficient}
\begin{figure}[h!]
\centering
\begin{subfigure}[t]{0.4\textwidth}
  \includegraphics[width=\textwidth]{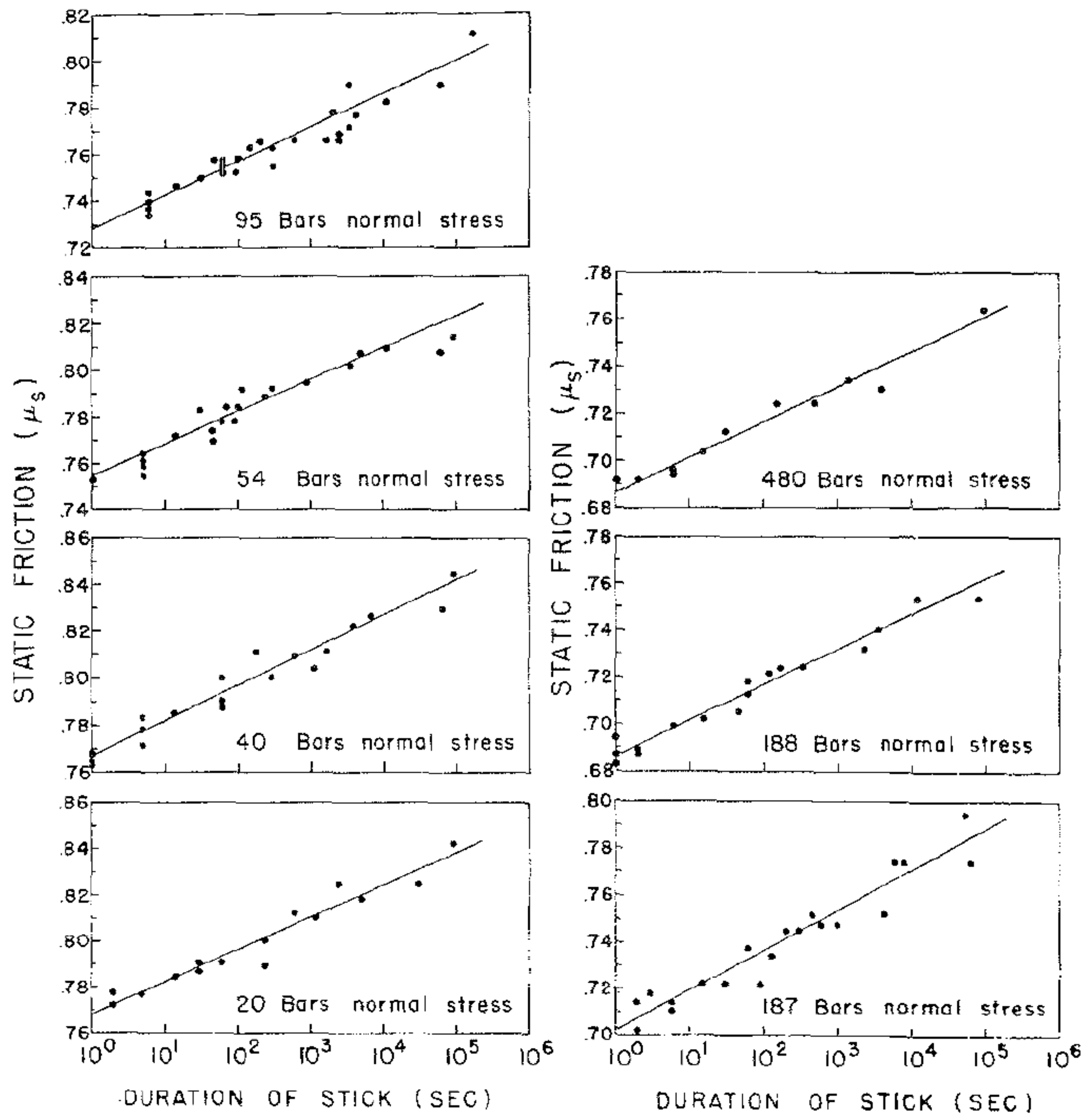}
\subcaption{}
\label{fig:static_friction_evolution_experimental_dieterich}
\end{subfigure} 
\begin{subfigure}[t]{0.5\textwidth}
  \includegraphics[width=\textwidth]{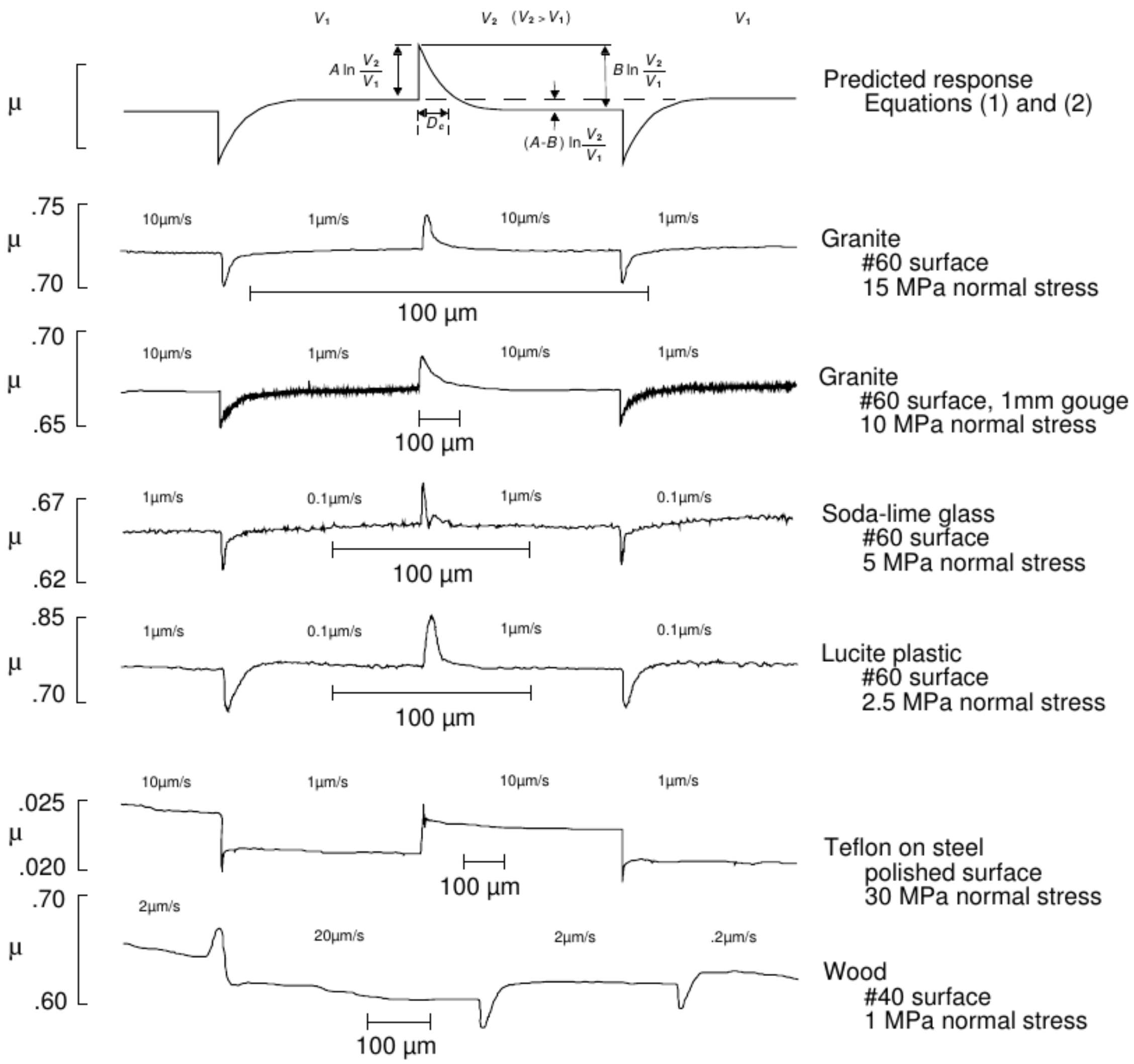}
\subcaption{}
\label{fig:vj_dieterich_1994}
\end{subfigure} 
\caption{(a) Evolution of the static friction coefficient with time at a constant 
         normal force for quartz sandstone. Different panels correspond to 
         different values of the normal force. Reproduced with permission from 
         \cite{dieterich:1}.
        (b) Evolution of kinetic friction coefficient in velocity jump experiments 
         on different materials. Reproduced with permission from  \cite{dieterich:3}.}
\end{figure}

Coulomb observed that the static friction coefficient $\mu_s$ is not a constant 
but depends on how long 
surfaces are in contact before sliding begins \cite{persson:1}. 
He proposed an empirical power law to fit his experimental results. Rabinowicz
\cite{rabinowiczE:1} and Dieterich \cite{dieterich:1}
 observed a similar  strengthening in their 
experiments on metals and rocks respectively. In Dieterich's experiments
\cite{dieterich:1}, 
rocks were held at constant normal and shear 
forces for varying lengths of time before the shear force was suddenly increased
to induce sliding. The coefficient of static friction was calculated as the 
ratio of the shear force required to initiate sliding to the applied normal 
force. $\mu_s$ was observed to increase logarithmically with the time of contact
(Figure \ref{fig:static_friction_evolution_experimental_dieterich}).
Dieterich proposed an empirical law to fit these results
\cite{dieterich:2}: 
\begin{equation}\label{eq:static_friction_evolution_experimental_dieterich}
 \mu_s(t) = \mu_0 + A \log (Bt+1),
\end{equation}
where $t$ is the time of 
stationary contact, $\mu_0, A$, and $B$ are constants dependent on the two surfaces 
across the interface. Typically, for rocks, $\mu_0$ is $0.7$-$0.8$, $A$ is 
$0.01$-$0.02$ and $B$ is of the order of $1$ second$^{-1}$ \cite{dieterich:2}. 

Two physical processes are conjectured to be the origin of the strengthening. 
First, because of high stresses, creep at contacts might result in
increased area of contact, leading to increased strength \cite{dieterich:3}.
Second, even if the area of contact does not change, the strength of each 
contact might increase with time \cite{bowdenFP:1}. 
\subsection{Kinetic friction coefficient}
Experiments have shown that the kinetic friction coefficient $\mu_k$ depends on
the sliding velocity. In a common type of experiments called velocity jump 
tests, two surfaces are slid at a constant velocity till the system reaches 
steady state. Then, a step change in sliding velocity is induced and the shear
force required to sustain this new velocity is monitored. Since the velocity is
constant, by equilibrium, this force must equal the frictional force at 
the interface. This shear force divided by the applied normal force is the 
kinetic friction coefficient. 
Experiments on different materials show \cite{dieterich:3} that, with a 
jump in velocity, $\mu_k$ also jumps (called direct effect), and the jump is 
followed by an evolution to a new steady state corresponding to the new sliding
velocity (Figure \ref{fig:vj_dieterich_1994}). 
The evolution happens over a characteristic length scale, with the time scale of
evolution to steady state being the ratio of the length scale to the sliding 
speed. This length scale, thought to be the slip 
necessary for the memory of the contacts to fade, is fairly independent of the 
sliding speed and the normal force but depends on the surface features 
\cite{dieterich:4}. Similar properties of the kinetic friction coefficient have
been found for sliding of thin granular layers \cite{marone:1}.

\subsection{Phenomenological laws}
A class of empirical laws called ``rate and state" (RS) laws has been proposed 
to capture the above experimental observations \cite{dieterich:4,ruina:1}. 
``Rate" here refers to the relative 
speed across the interface and ``state" refers to  one or more internal
variables used to represent the memory in the system. These laws,
used widely in simulations of earthquake phenomena, have been successful in
reproducing many of the observed features of earthquakes
\cite{tse:1,dieterich:9,kaneko:1,barbot:1,noda:1}. One commonly used RS law 
with a single state variable takes the form: 
\begin{equation}\label{eq:RS}  
 \mu_k = \mu_0 + a \ln (\frac{v}{v^*}) + b \ln (\frac{v^*\theta}{D_c}),
\end{equation}
where $\mu_k$ is the coefficient of friction, $v$ is the sliding velocity, $a,b,v^*,
\mu_0$, and $D_c$ are constants, and $\theta$ is an internal variable with 
dimensions of time. An evolution law is prescribed for the internal variable 
$\theta$. Two well-known laws are the aging law,
\begin{equation}\label{eq:aging_law}  
\dot{\theta} = 1 - \frac{v \theta}{D_c},
\end{equation}
and the slip law,
\begin{equation}\label{eq:slip_law}  
\dot{\theta} =  - \frac{v \theta}{D_c} \ln(\frac{v \theta}{D_c}).
\end{equation}
In these equations, $D_c$ is the characteristic length scale over which the 
coefficient
of friction evolves to its steady state in the jump test. 
$D_c$ is related to the roughness of the sliding surfaces and is of the order of
microns for most engineering surfaces \cite{dieterich:4,dieterich:8}.
At steady state, $\dot{\theta} = 0$, and from Equation (\ref{eq:aging_law}) or 
(\ref{eq:slip_law}), $\theta_{ss}(v) = D_c/v$. Using this in Equation 
(\ref{eq:RS}), the steady state friction coefficient is given by,
\begin{equation}\label{eq:rate_state_steady_state}
\mu_{ss}(v) = const + (a-b) \ln(\frac{v}{v^*}).
\end{equation}
If $a-b>0$, the steady state friction coefficient increases with increasing 
sliding speed and if $a-b<0$, the steady state friction coefficient decreases 
with increasing sliding speed. The two cases are known as velocity-strengthening 
and velocity-weakening respectively. This has implications, for example for the
stability of sliding, a requirement for stick-slip being $a-b<0$ 
\cite{ruina:1,ruina:2}.

\section{Collective behavior of viscoelastic asperities}
\label{sec:Viscoelastic asperity model}

\subsection{Basic elements}

The basic ingredients of our model are a constitutive description of single 
asperities at the microscale, and a stochastic characterization of rough 
surfaces. We now describe these in detail.

\subsubsection{Single asperity}
\label{subsuc:singleAsperity}
The behavior of an asperity depends on various factors such as material 
properties, local stresses, sliding speed etc.
As a first step, we model
asperities as being viscoelastic, using a spring-dashpot system known
as a Standard Linear Solid (SLS, see Figure \ref{fig:sls_system}).
An SLS consists of a spring in parallel with a spring and dashpot in series.
The equation for the evolution of the force $F$ on an SLS as a function of 
its length and the rate of change of its length is:
\begin{equation}\label{eq:sls_force_evolution} 
\dot{F} = (k_1+k_2)\dot{x} - \frac{k_2}{\eta} F + \frac{k_1 k_2}{\eta}(x-x^0),
\end{equation}
where $k_1, k_2$ are the stiffnesses of springs 1 and 2 respectively, 
$\eta$ is the viscosity of the dashpot, $x, x^0$ are the current and 
undeformed lengths of spring 1 and dot denotes the time derivative. An asperity
can be in two states, in contact or out of contact with the surface it slides 
on. When in contact, 
its length evolution is known and the force evolution can be calculated using 
Equation (\ref{eq:sls_force_evolution}). When out of contact, the force on the
asperity is zero and, setting $F=0$ and $\dot{F} = 0$ in the 
Equation (\ref{eq:sls_force_evolution}), the evolution of its length is given by:
\begin{equation}\label{eq:sls_relaxation} 
\dot{x} = - \frac{k_1 k_2}{\eta(k_1+k_2)} (x-x^0).
\end{equation}

A natural question to ask is whether we can relate the material 
and geometric properties of the asperities to the parameters $k_1, k_2$ and 
$\eta$ of the SLS. One way to do this would be to solve a viscoelastic 
Hertzian problem. Assuming that the asperity is spherical and the material is
linear viscoelastic, from the solution of the elastic 
Hertzian problem, the viscoelastic Hertzian problem can be solved using 
the method of Laplace transforms \cite{chengL:1}. The contact, initially at 
zero force and deformation, is instantaneously brought to a deformation 
$\delta_0$ and the evolution of the force
is computed. The instantaneous force is related to the instantaneous stiffness 
of the SLS, $k_1+k_2$, the steady state force is related to $k_1$, and the 
rate of relaxation to steady state is related to $\eta$. The 
Hertzian contact problem is nonlinear whereas the SLS element is
linear. The nonlinearity of the Hertzian problem manifests as the dependence of
$k_1,k_2$, and $\eta$ on the deformation $\delta_0$. We can however, get an 
order-of-magnitude estimate of the values of the parameters in our model.
Results from such calculations for two materials, Polyvinyl Alcohol (PVOH) 
and Polystyrene \cite{chengL:1}, are given below. 

PVOH: $$ \frac{k_1}{\sqrt{\rho \delta_0}} = 0.18 \text{ GPa}, 
         \frac{k_2}{\sqrt{\rho\delta_0}} = 0.15 \text{ GPa},
         \frac{\eta}{\sqrt{\rho \delta_0}} = 0.48 \text{ GPa-s}.$$

Polystyrene: $$ \frac{k_1}{\sqrt{\rho \delta_0}} = 2.75 \text{ GPa}, 
                \frac{k_2}{\sqrt{\rho \delta_0}} = 0.39 \text{ GPa},
                \frac{\eta}{\sqrt{\rho \delta_0}} = 4.51 \text{ GPa-s}.$$
Above, $\rho$ is the radius of curvature of the asperity. 

\subsubsection{Rough surfaces}
Rough surfaces have been characterized by representing the 
heights from a reference level as a stochastic process \cite{longuethiggins:1,
saylesRS:1}.  
This characterization has been used extensively in exploring various aspects of
contact between surfaces \cite{greenwood:1,nayakPR:1,nayakPR:2}.
Profile measurements have shown that many types of surfaces can be 
modeled as a Gaussian noise with an exponential correlation  \cite{whitehouse:1}.
Such a noise, known as an Ornstein-Uhlenbeck process, satisfies the following 
stochastic differential equation:
\begin{equation}\label{eq:OU_noise}
\frac{dy}{dz} = -\frac{1}{\lambda} y(z) + \sqrt{\frac{2}{\lambda}} \sigma \zeta(z).
\end{equation}
Here, $y$ is the height of the surface from a reference level (chosen such that 
the mean height is zero), $z$ is the horizontal spatial variable, 
$\lambda$ is the correlation length, 
$\sigma$ is the rms-roughness of the surface and $\zeta(z)$ 
is a Gaussian white noise with unit standard deviation. For typical surfaces,
$\sigma$ and $\lambda$ are of the order of a few microns \cite{whitehouse:1}. 
We adopt this description of a rough surface. We also 
assume that the surface is rigid since this considerably simplifies our 
calculations. 

\subsubsection{Local friction law}
\label{subsec:localFrictionLaw}
To determine the coefficient of friction, we need the macroscopic 
normal and shear forces. To compute the macroscopic shear force as a moment with
respect to the probability distribution of microscopic normal forces, we need 
to know how the normal and shear forces are related for a single asperity. 
We assume local friction laws of the form,
\begin{equation}\label{eq:localFrictionLaw}
s(F,v) =  f(v) F + c_2 F^n, \quad f(v) =  \left\{ \begin{array}{ll}
  0 &\mbox{$v \leq v_c$} \\
  \mu_0 + c_1 \log(\bar{v}) &\mbox{$v>v_c$.}
       \end{array} \right.
\end{equation}
$s$ here is the shear force at the contact, $F$ is the local
normal force, $v$ is the sliding speed, $\bar{v}$ is the nondimensional
sliding speed,
$\mu_0,n, c_1$, and $c_2$ are constants, $v_c$ is the cutoff velocity for the
velocity dependence of friction, and $2/3 \leq n \leq 1$.

The local friction law depends on the material and geometry of
the contacts. If a contact is elastic and its geometry is 
spherical, then, by the Hertz theory of contact, the area of contact 
varies as the two-thirds power of the normal force. Further, if 
the contact has a shear strength $\tau_{max}$, then the local friction law 
is: 
$$ A \propto F^{2/3}, \qquad s = \tau_{max}A \propto F^{2/3}. $$ 
As in the Bowden and Tabor model, if contacts are 
plastic because of the high local stresses, the area of contact is 
proportional to the normal force and the local friction law is:
$$ A \propto F^{1}, \qquad s = \tau_{max}A \propto F^{1}. $$ 
An actual contact may be in between these two limiting cases and 
thus, the power in the local friction law between $2/3$ and $1$. Also, 
different contacts in the population may be in different states.

If the surfaces are sliding at a relative speed $v$, the 
asperities in contact are sheared at a strain rate proportional to the sliding
speed, and if the shear resistance depends on the strain rate, then the local
friction law will be velocity-dependent. Taking cue
from experimental results, we assume this velocity dependence to be 
logarithmic. A theoretical justification for the logarithmic dependence has been
proposed by Rice et al \cite{rice:1}.

Alternatively, local friction laws can be derived from theoretical and 
experimental studies of single asperity contacts. For example, Kogut and Etsion
\cite{kogutL:1} consider the inception of sliding of a single spherical 
elastoplastic contact and conclude that $\mu \propto F^{-0.345}$
when the normal force by itself does not cause any plastic deformation 
($\mu$ here is the single asperity friction coefficient). Note
that this is very close to the Hertzian $s \propto F^{2/3} $
approximation above. Archard \cite{archard:2}, using a crossed cylinder 
apparatus, reports $\mu \propto F^{-0.26}$ 
for perspex and $\mu \propto F^{-1/3}$ for brass.
Wandersman et al \cite{wandersmanE:1}, looking at texture-induced  modulations
of friction, report that $s \propto F^{0.87}$ for an elastomer on glass.\medskip
\medskip

To summarize, our model for two rough surfaces in contact consists of an 
ensemble of independent viscoelastic SLS elements sitting on a rigid rough 
surface modeled as an Ornstein-Uhlenbeck process. 
With this model, we simulate the static contact and velocity jump experiments
to study the time and velocity dependence of friction.

\subsection{Static friction}

\subsubsection{Formulation}
In this section, we study the evolution of an ensemble of independent 
SLS elements in static contact with a rigid rough Gaussian surface 
(Figure \ref{fig:sls_system}) under a global normal force $F_N$. 
The evolution of each SLS is governed by Equations (\ref{eq:sls_force_evolution}) 
and (\ref{eq:sls_relaxation}).
For simplicity, we assume that the values of $k_1,k_2,\eta$
and $x^0$ are the same for every element in 
the ensemble. Without loss of generality, we can 
set $x^0 = 0$ since this corresponds to choosing a particular reference level to 
measure the length $x$. Dilatation, which 
is the distance between the reference planes from which the lengths of the SLS
elements $x$ and the heights of the rigid surface $y$ are measured, is denoted
by $d$. 
At any instant $t$ during the evolution of the system, only a fraction of the 
elements are in contact with the surface and for these, the contact condition
implies:
\begin{equation}
x(t)+y=d(t). 
\end{equation}
Since we have assumed the parameters
$k_1,k_2,\eta,x^0$ to be the same for all the elements in the ensemble, each 
element can be labeled by the height $y$ of the rigid surface that is sees.
At time $t$, the global normal force is given by: 
\begin{equation}\label{eq:sc_normal_force_moment_y}
F_N(t) = \mathbb{E}_y(F(t,y)) = \int_{d(t)}^\infty F(t,y)P_y(y)\mathrm{d}y,
\end{equation}
where $F(t,y)$ is the force on an SLS corresponding to time $t$ and height $y$ 
of the rough surface, and $P_y(y)$ is the probability distribution of heights of
the rough surface. The limits of integration are from $d(t)$ to $\infty$ since
elements are in contact with the surface only if the height $y$ is greater 
than $d(t)$ (since $x^0=0$).
For a Gaussian distribution of surface heights, we have,
\begin{equation}
P_y(y)  = \frac{1}{\sqrt{2\pi}\sigma} e^{-\frac{y^2}{2 \sigma^2}},
\end{equation}
where $\sigma$ is the rms-roughness of the surface. 
The elements being viscoelastic, the 
force in the ones in contact would decrease with time if the dilatation were 
constant. To reproduce experimental conditions, we constrain the total normal
force to be constant. 
To satisfy this constraint, dilatation has to decrease with time. 
Differentiating Equation (\ref{eq:sc_normal_force_moment_y}) with respect to 
time gives: 
\begin{equation}\label{eq:sc_normal_force_derivative}
\dot{F}_N(t) = \int_{d(t)}^\infty \frac{\partial{F}}{\partial{t}} (t,y)P_y(y)
               \mathrm{d}y - F(t,d) P_y(d) \dot{d}(t).
\end{equation}
$F(t,d)=0$ since $y=d(t)$ implies that the contact is formed at 
time $t$. The evolution equation for the force on an SLS (Equation 
(\ref{eq:sls_force_evolution}) with $\dot{x}=\dot{d}$; $\dot{y}=0$ as the 
surface is rigid) gives: 
\begin{equation}\label{eq:sc_sls_evolution_force}
\frac{\partial{F}}{\partial{t}} (t,y) = (k_1+k_2) \dot{d}(t) - 
                   \frac{k_2}{\eta} F(t,y) + \frac{k_1 k_2}{\eta} (d(t)-y).
\end{equation}
Using $F(t,d)=0$ and Equation (\ref{eq:sc_sls_evolution_force})  
in Equation (\ref{eq:sc_normal_force_derivative}), the evolution equation for 
the dilatation at constant normal force  ($\dot{F}_N=0$) is: 
\begin{equation}\label{eq:sc_dilatation_evolution}
\dot{d}(t) = \frac{\frac{k_2}{\eta}F_N - \frac{k_1 k_2}{\eta}\int_{d(t)}
             ^\infty (d(t)-y) P_y(y) \mathrm{d}y}{(k_1+k_2)\int_{d(t)}^\infty 
             P_y(y) \mathrm{d}y}. 
\end{equation}
We now turn to the static friction force.
Knowing the local friction law, $s = s(F)$, we can determine the global shear
force as, 
\begin{equation}\label{eq:sc_shear_force_moment_y}
F_S(t) = \mathbb{E}_y(s(F(t,y))) = \int_{d(t)}^\infty s(F(t,y))P_y(y)\mathrm{d}y,
\end{equation}
and the coefficient of friction can be calculated as the ratio of the two 
forces, 
\begin{equation} 
 \mu_s(t) = \frac{F_S(t)}{F_N}.
\end{equation}
Before proceeding further, we nondimensionalize the equations.

\subsubsection{Nondimensionalization}
To nondimensionalize time, we use $\eta/k_2$, which is one of the two
characteristic timescales 
of the SLS. A natural length scale in the problem is the rms-roughness of the 
surface, $\sigma$. For forces, we use $k_1 \sigma$. Using these characteristic
quantities, we define the following nondimensional variables: 
\begin{equation}
   \bar{t} = \frac{t}{\eta/k_2}, \quad \bar{d} = \frac{d}{\sigma},
   \quad \bar{y} = \frac{y}{\sigma},\quad  \bar{F} = \frac{F}{k_1 \sigma}.
\end{equation}
Equation (\ref{eq:sc_sls_evolution_force}) after nondimensionalization is:
\begin{equation}\label{eq:sc_sls_evolution_force_nondim}
\frac{\partial{\bar{F}}}{\partial{\bar{t}}} (\bar{t},\bar{y}) = 
R \bar{d}' - \bar{F}(\bar{t},\bar{y}) + \bar{d}(\bar{t})-\bar{y}.
\end{equation}
The nondimensional equation for the dilatation evolution is:
\begin{equation}\label{eq:sc_dilatation_evolution_nondim}
\bar{d}' = \frac{\bar{F}_N(\bar{t}) - \int_{\bar{d}}
             ^\infty (\bar{d}-\bar{y}) P_{\bar{y}}(\bar{y}) \mathrm{d}\bar{y}}
       {R\int_{\bar{d}}^\infty P_{\bar{y}}(\bar{y})\mathrm{d}\bar{y}},
\end{equation}
and the coefficient of friction is given by:
$$ \mu_s(\bar{t}) = \frac{\bar{F}_S(\bar{t})}{ \bar{F}_N}.$$
In the above equations, prime denotes differentiation with respect to 
nondimensionalized time, 
\begin{equation}
R = 1+\frac{k_2}{k_1}, 
\end{equation}
and the  probability distribution of the normalized surface heights
$P_{\bar{y}}(\bar{y})  = 
\frac{1}{\sqrt{2\pi}} e^{-\frac{\bar{y}^2}{2}}$.

After nondimensionalization, we have two parameters: $R$, which is the 
ratio between the instantaneous and steady state stiffnesses, and $n$, the 
power in the local friction law.  The local friction exponent $n$ is indicative 
of the state of the contact, being $2/3$ for elastic and $1$
for plastic contact.

\subsubsection{Static friction evolution}
Suppose that, at $t=0$, an ensemble of SLS elements is instantaneously brought
to a dilatation $d=d_0$ against the rigid surface. Some of the SLS elements come
into contact with the surface and result in a total normal force $F_N$. The 
system is then allowed to evolve at this constant global normal force $F_N$. The
dilatation evolves to satisfy the constraint of constant $F_N$. This evolution
can be calculated using Equation (\ref{eq:sc_dilatation_evolution_nondim}). 
From the dilatation history, the evolution of the probability density of the
normal forces can be determined using 
Equation (\ref{eq:sc_sls_evolution_force_nondim}).  Using this and the local
friction law, we can determine the evolution of the static friction coefficient.

\begin{figure}[h!]
\begin{center}
\includegraphics[scale=0.5]{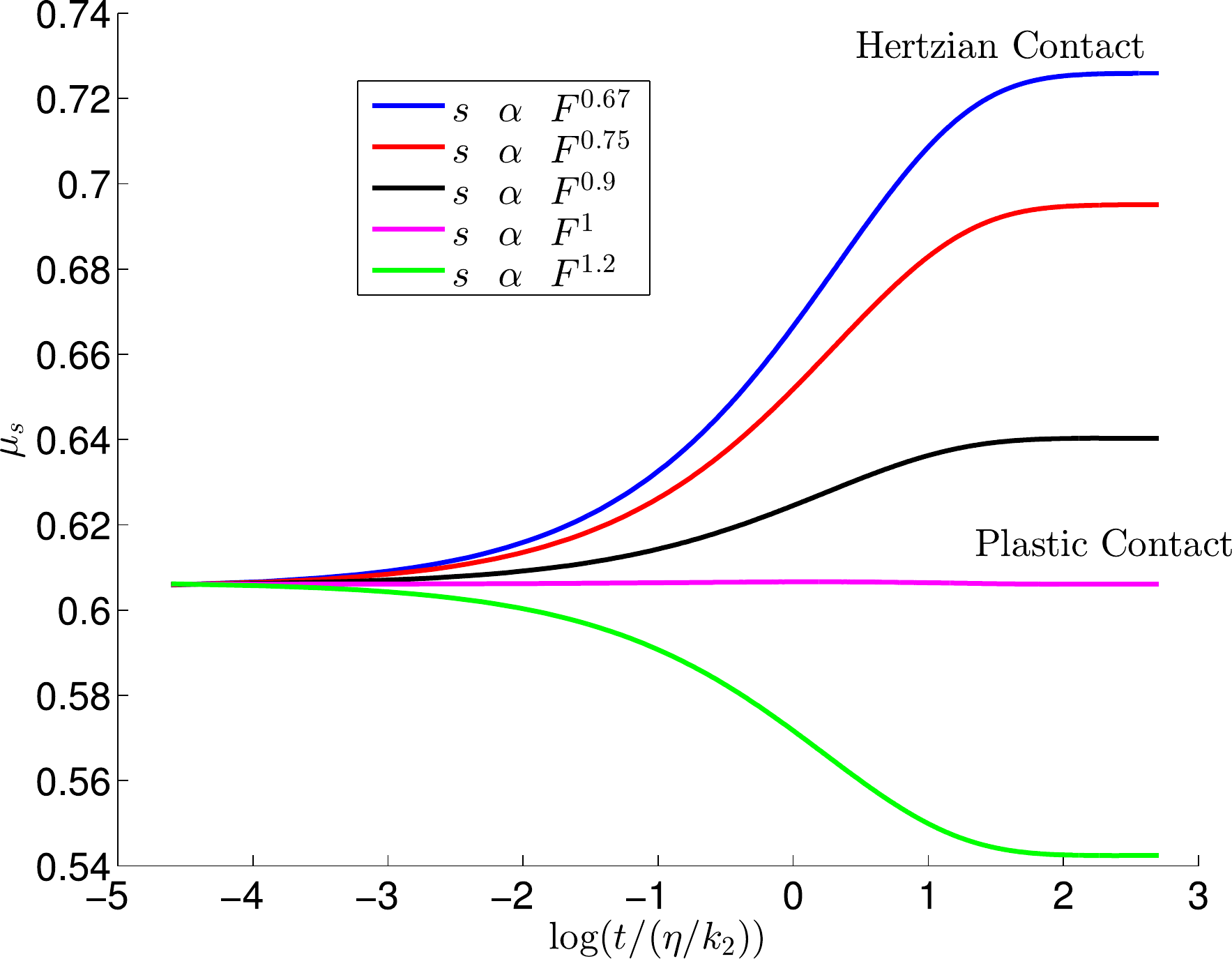}
\caption{Evolution of the static friction coefficient ($\mu_s$) with time at a constant global
normal force for different local friction exponents $n$.  For the physically relevant
values of $n \in [2/3, 1)$, $\mu_s$ increases logarithmically with time 
for a range of times around the relaxation time of each asperity.
For the unphysical case $n>1$, $\mu_s$ decreases with time, a behavior that has
not been observed in experiments. }
\label{fig:frictionEvolutionStaticContactLocalFrictionLaw}
\end{center}
\end{figure}

Figure \ref{fig:frictionEvolutionStaticContactLocalFrictionLaw} shows the evolution of
$\mu_s$ for different local friction exponents. 
At small times, $\frac{t}{\eta/k_2} \ll 1$, $\mu_s$ remains constant since the
SLS elements need a finite
time to start relaxing. $\mu_s$ then starts increasing and, around
the relaxation time of the SLS, $\frac{t}{\eta/k_2} = \mathcal{O}(1)$, the 
increase is approximately logarithmic in time.  
At large times,
$\frac{t}{\eta/k_2} \gg 1$, all the SLS elements have relaxed to their steady 
state and $\mu_s$ evolves to a constant value. 
The variation in $\mu_s$ is largest for the case of elastic contacts
($n=2/3$). As the power
in the local friction law reaches $1$, the case of plastic contacts, we recover
the result of the Bowden and Tabor model in which the friction coefficient is 
a constant given by the ratio of the shear strength to the hardness of the 
contacts. For $n>1$, $\mu_s$ decreases with the time of contact. As explained 
in section \ref{subsec:localFrictionLaw}, we expect $n\in[2/3,1]$ for real
materials.  We believe this to be the reason why we only see an increase
of $\mu_s$ in experiments.
\begin{figure}[h!]
\centering
\begin{subfigure}[t]{0.4\textwidth}
\includegraphics[width=\textwidth]{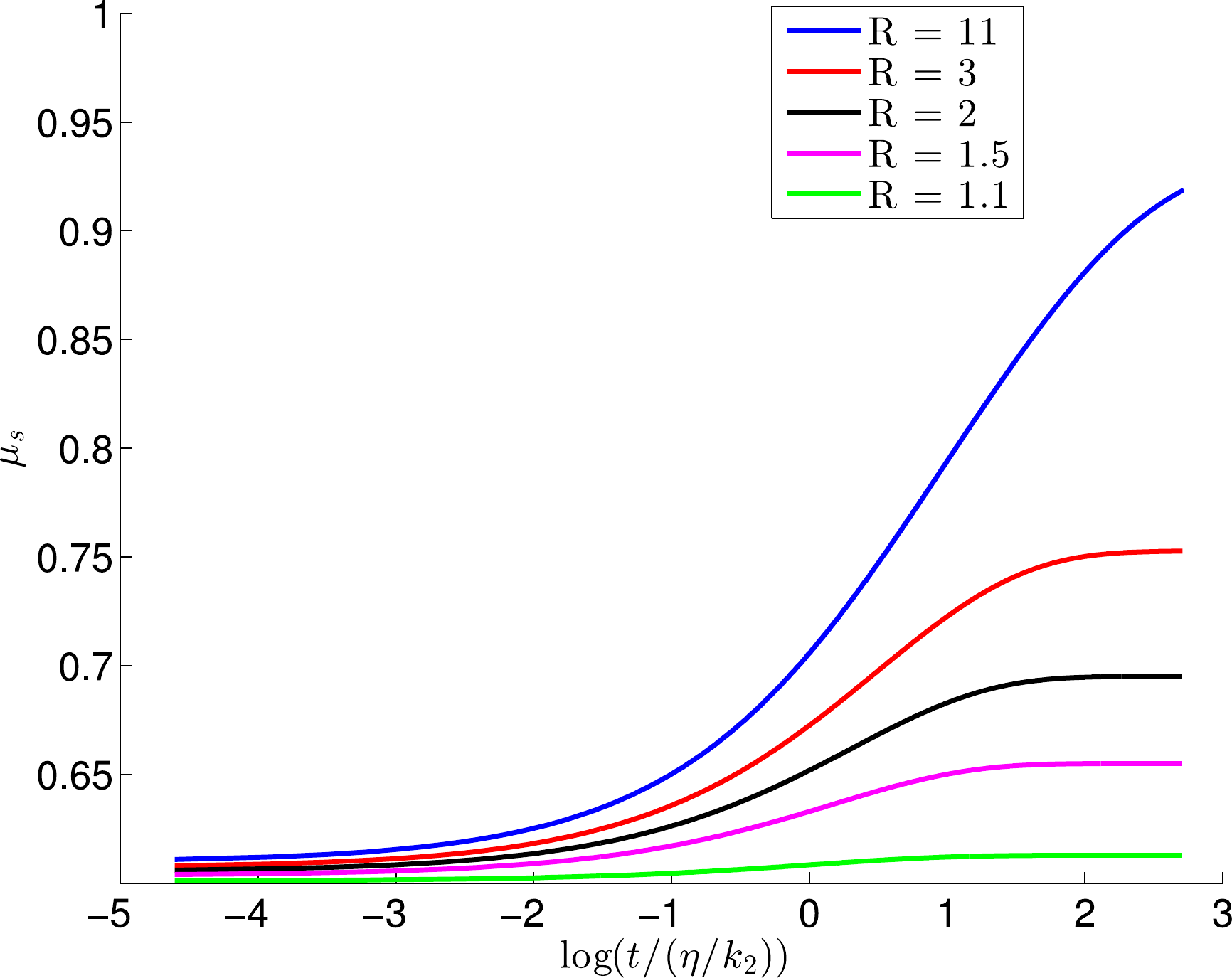}
\subcaption{}
\label{fig:frictionEvolutionStaticContactStiffnessRatio}
\end{subfigure} 
\quad
\begin{subfigure}[t]{0.4\textwidth}
\includegraphics[width=\textwidth]{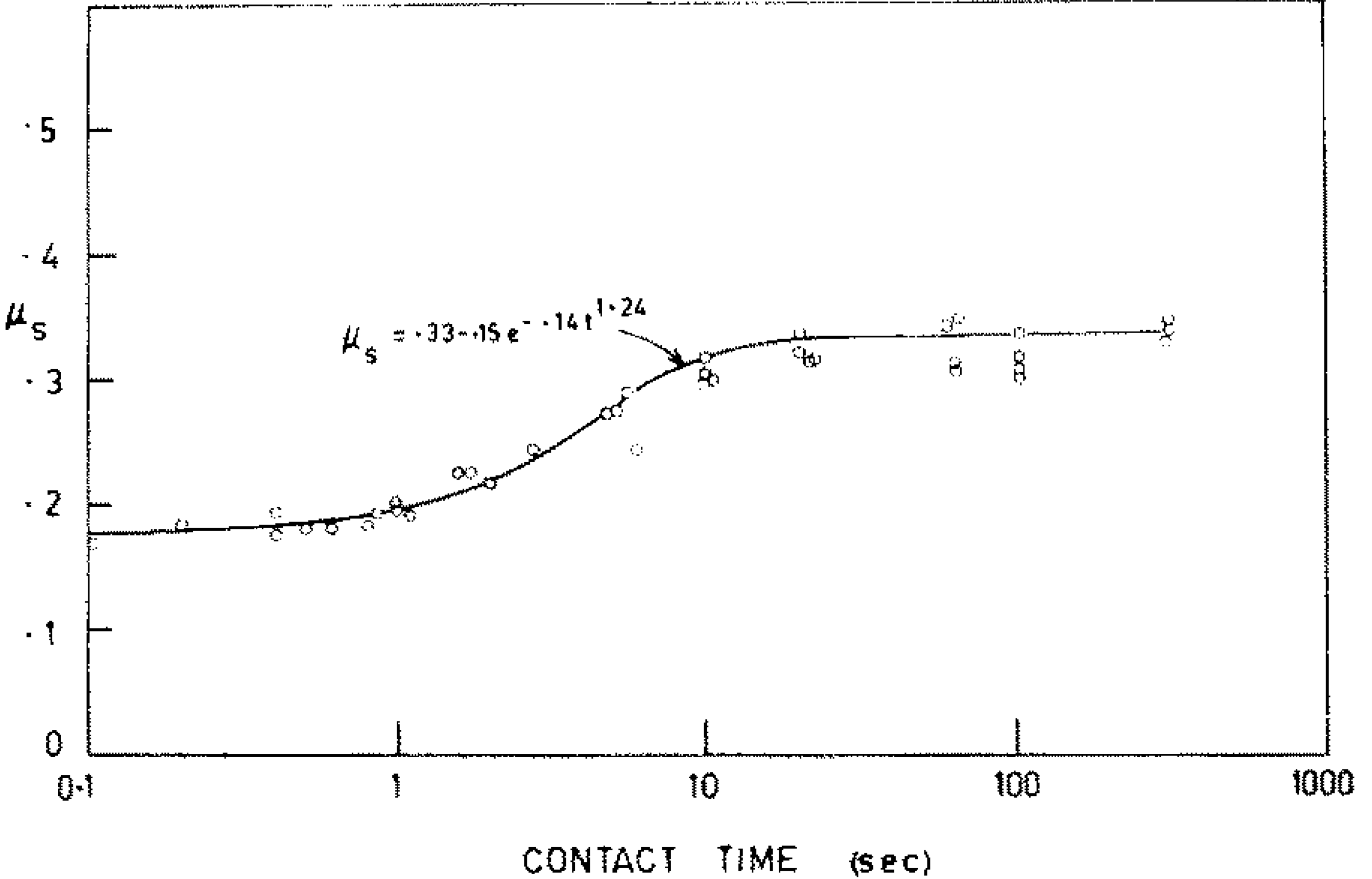}
\subcaption{}
\label{fig:frictionEvolutionExperimentalRichardsonTolle}
\end{subfigure} 
\caption{ (a) Evolution of friction coefficient with time at a constant global 
         normal force for different stiffness ratios, $R=1+k_2/k_1$. 
 The system reaches steady state faster for smaller $R$.
         (b) The saturation at small and large times has also been observed in 
 some experiments (reproduced with permission from \cite{richardsonRSH:1}).}
\end{figure}

Figure \ref{fig:frictionEvolutionStaticContactStiffnessRatio} shows the evolution of
$\mu_s$ for different values of $R$, using a local friction law 
$s(F) \propto F^{0.75}$. When $R \approx 1$, or $k_2\ll k_1$, the instantaneous
and steady state stiffnesses 
are not very different and hence, the growth in the friction coefficient is 
small. For $k_2\gg k_1$, the instantaneous and steady state stiffnesses are very
different and consequently, we see a larger growth in $\mu_s$. Also, 
Figure \ref{fig:frictionEvolutionStaticContactStiffnessRatio} shows that the 
steady state is reached faster when $R$ is smaller. The SLS 
has two characteristic timescales, $\eta/k_2$ and $\frac{\eta}{k_2}(1+k_2/k_1)$.
Since we have nondimensionalized time with $\eta/k_2$, $R$ is the second 
characteristic timescale in nondimensionalized time and thus, for smaller $R$, the 
steady state is achieved faster. 

The predicted increase of the coefficient of static friction with time has been
widely observed in experiments \cite{rabinowiczE:1,dieterich:1,richardsonRSH:1}.
If $\eta/k_2$ is of $\mathcal{O}(1)$ seconds, as it is for the two materials 
mentioned earlier, we see that the friction coefficient reaches its steady 
state at around $10^2$ seconds and the total variation in $\mu_s$ lasts about 
$3$-$4$ decades in time. 
In some materials like mild steel, the predicted behavior -- linear 
increase with logarithmic time over a few decades with a lower and upper 
saturation for small and large times -- is also in agreement with experimental
observations \cite{odenJT:1,richardsonRSH:1}, as shown in the Figure
\ref{fig:frictionEvolutionExperimentalRichardsonTolle}.
In the experiments on rocks \cite{dieterich:1,dieterich:2} however, the logarithmic
growth persists through the duration of the experiment (some experiments have
lasted six decades in time).  Since $\mu_s$ cannot increase indefinitely, it 
eventually has to reach a steady state. This delayed saturation is not
captured by our model.  We conjecture that the difference 
between our model response and the rock experiments could
be for the following reasons. The SLS element has only one relaxation time,
while a 
real viscoelastic material has many relaxation timescales.
In Figures \ref{fig:frictionEvolutionStaticContactLocalFrictionLaw} and 
\ref{fig:frictionEvolutionStaticContactStiffnessRatio}, the x-axis is the 
nondimensionalized logarithmic time. When plotted with respect to logarithmic
time, changing $\eta/k_2$ corresponds to translating the curve horizontally
by $\log(\eta/k_2)$. Thus, in the case of multiple relaxation timescales 
(multiple $\eta/k_2$), the region of $\mu_s$ increase will be wider.
Furthermore, if
the asperities interact with each other through the bulk, the interactions can 
result in a continuum of timescales for the response at the macroscopic scale.
Exploring this hypothesis remains a topic of current work.

\subsection{Kinetic friction}
\subsubsection{Formulation}
Consider a single SLS sliding on a rough rigid surface 
(Figure \ref{fig:sls_system}). The slider can be in two states, in contact or 
out of contact with the surface. When in contact, its length evolves to conform
to the rigid surface, $x(t)=d(t)-y(t)$ and the force evolves according to 
Equation (\ref{eq:sls_force_evolution}). Since we have modeled 
the rough surface as a stochastic process, the differential equation for 
force evolution during contact will also be a stochastic one. 
If the SLS ensemble slides at a constant velocity $v$, then the horizontal 
coordinate is $z =vt$. Using chain rule, we can change the independent variable 
from the horizontal coordinate $z$ to time $t$ in Equation (\ref{eq:OU_noise})
for the rough surface: 
\begin{equation}\label{eq:OU_noise_time}
\frac{dy}{dt} =\frac{dy}{dz}\frac{dz}{dt} = 
v \frac{dy}{dz} = 
-\frac{v}{\lambda} y(t) + \sqrt{\frac{2v}{\lambda}} \sigma \zeta(t).
\end{equation}
When the slider is in contact with the surface, $\dot{x} = \dot{d}-\dot{y}$.
Using this and Equation (\ref{eq:OU_noise_time}) in Equation 
(\ref{eq:sls_force_evolution}), the stochastic differential 
equation for the evolution of the normal force during contact is:
\begin{equation}\label{eq:force_langevin}
\dot{F}(t) =  -\frac{k_2}{\eta} F(t) + (k_1+k_2)\dot{d}(t) +\frac{k_1 k_2}{\eta}
               (d(t)-x^0) + \left(\frac{(k_1+k_2)v}{\lambda} - \frac{k_1 k_2}{\eta}
               \right) y(t) - (k_1+k_2)\sqrt{\frac{2v}{\lambda}}\sigma \zeta(t).
\end{equation}
Here, the terms $y(t)$ and $\zeta(t)$ are the Ornstein-Uhlenbeck and white 
noise, respectively. Their statistical properties are:
\begin{equation}\label{eq:noise_cross_correlation}
<y(t)> =0, \qquad <y(t_1)y(t_2)> = \sigma^2 e^{-\frac{v|t_1-t_2|}{\lambda}}, 
\end{equation} 
$$<\zeta(t)> = 0, \qquad <\zeta(t_1) \zeta(t_2)> = \delta(t_1-t_2), $$	
$$ <\zeta(t_1)y(t_2)> = \begin{cases} 0, & \mbox{if } t_1 > t_2, \\ 
                      \sqrt{\frac{2v}{\lambda}}\sigma e^{-v|t_2-t_1|/\lambda}, 
                      & \mbox{if } t_1 <= t_2, \end{cases}$$ 
where $<>$ denotes ensemble average. Since $y(t)$ is driven by $\zeta(t)$ 
(Equation (\ref{eq:OU_noise_time})), the cross correlation between the two noises
is not zero. 

When the slider is not in contact, the normal force is zero and its length 
evolves 
according to Equation (\ref{eq:sls_relaxation}). Before proceeding further, we 
nondimensionalize the above system of equations.

\subsubsection{Nondimensionalization}
As before, we use $\sigma$ to nondimensionalize length, $\eta/k_2$ to 
nondimensionalize time and $k_1 \sigma$ to nondimensionalize force, 
and use a bar to denote non-dimensional quantities.
Equation (\ref{eq:force_langevin}) for an SLS in contact, after 
nondimensionalization is:
\begin{equation} \label{eq:force_langevin_nondim}
\bar{F}' = -  \bar{F} + R \bar{d}' + \bar{d}-\bar{x^0}+
             \left(\frac{R\bar{v}}{\bar{\lambda}}-1\right)\bar{y}-R\sqrt{\frac{2\bar{v}}{\bar{\lambda}}}\bar{\zeta},
\end{equation}
\begin{equation}\label{eq:noise_statistics_nondim}
<\bar{y}> = 0, \quad <\bar{y}(\bar{t}_1)\bar{y}(\bar{t}_2)> = 
                  e^{-\bar{v}|\bar{t}_1-\bar{t}_2|/\bar{\lambda}}, \\
\end{equation}
$$ <\bar{\zeta}> = 0, \quad <\bar{\zeta}(\bar{t}_1)\bar{\zeta}(\bar{t}_2)> = 
                  \delta(\bar{t}_1-\bar{t}_2).$$
After nondimensionalization, Equation (\ref{eq:sls_relaxation}) for an
SLS out of contact becomes:
\begin{equation}\label{eq:sls_relaxation_nondim}
 \bar{x}' = -\frac{1}{R}(\bar{x}-\bar{x}^0).
\end{equation}
As before prime denotes differentiation with respect to $\bar{t}$. 

$\bar{x}^0$, the undeformed length of the viscoelastic
sliders, may be set  to $0$ since this is equivalent to choosing a reference 
level. $\bar{\lambda}$ is the correlation length of the surface, which we set 
to $1$. In the governing equations, $\bar{\lambda}$ appears only as $\bar{v}/
\bar{\lambda}$, thus $\bar{\lambda}$ sets a scale for the sliding speed.

This leaves the following non-trivial parameters in the model.  
$R$, as in the static contact case, is the ratio of the instantaneous and 
steady state stiffnesses of the SLS.
$\bar{v} = k_2v/\eta\sigma$, a nondimensional sliding speed,
is the number of rms-roughness lengths that the SLS slides in one 
relaxation time $\eta/k_2$. 
If $\bar{v}\gg1$, the SLS has little time to relax when in contact 
and hence its response will be 
close to its instantaneous elastic response. If $\bar{v}\ll1$, the SLS has time
to relax to its steady state and its response will be similar to its steady 
state elastic response.  

We have four parameters in the local friction law, 
$\mu_0, c_1, c_2, n$ (Equation \ref{eq:localFrictionLaw}).
$\mu_0$ sets a reference value for the coefficient of friction, and we choose 
this to be $0.6$.  The constant $c_1$ controls velocity dependence of the 
local friction and it is set to  $0.01$ unless otherwise mentioned.
The constant $c_2$ describes the evolution of the friction on steady sliding,
and it is set to $0.2$.
Unless mentioned otherwise, the local friction exponent $n$ is set to
$0.67$ since the transient is most pronounced in this case.

\subsubsection{From one to many}
As the surfaces slide, each asperity sees a different profile. 
If two asperities are close to each other,
the profiles they see will be correlated. We neglect this correlation and 
associate an independent realization of the noise $\bar{y}(\bar{t})$ as the
profile on which an SLS element slides. The 
question we seek to answer is: Given the statistical properties of
$\bar{y}(\bar{t})$, what are the statistical properties of
the force $\bar{F}$? In particular, what is the probability density of force
$P(\bar{F})$? 
A stochastic equation such as Equation (\ref{eq:force_langevin_nondim}) is known 
as a Langevin equation. Averaging this equation over the ensemble of realizations
of the noise $\bar{y}(\bar{t})$, one can derive a partial differential equation for the 
evolution of the probability density $P(\bar{t},\bar{F})$ \cite{riskenH:1}. 
An example of this is the
heat equation which results from the averaging of the stochastic equation 
corresponding to the Brownian motion of a single particle. We would like to do a
similar ensemble averaging of Equation (\ref{eq:force_langevin_nondim}). 
The problem, however, is that 
we have two sources of noise, $\bar{y}(\bar{t})$ and $\bar{\zeta}(\bar{t})$, 
and the two are correlated (Equation (\ref{eq:noise_cross_correlation})). 
In such a 
case, there is no known method of deriving the partial differential equation
for the evolution of probability density. Hence, we resort to a numerical Monte
Carlo method where we generate an ensemble of sliders and surface profile
realizations, evolve the system at the microscale, and compute statistics of the
ensemble to determine macroscopic properties. There is however, a particular
case of pure white noise, where we can derive a partial differential
equation for the evolution of $P(\bar{t},\bar{F})$.

\subsubsection{The case of pure white noise}
\label{subsec:FokkerPlanck}
In Equation (\ref{eq:force_langevin_nondim}), if the velocity 
$\bar{v} = \bar{\lambda}/R$, the coefficient of $\bar{y}$ becomes zero.
We are then left with only the white noise term and we can derive a partial
differential equation (Fokker-Planck equation) for the evolution of $P(\bar{t},F)$ . 
The Fokker-Planck equation corresponding to the Langevin equation 
(\ref{eq:force_langevin_nondim}) when $ \bar{v} = \bar{\lambda}/R$ is 
\cite{riskenH:1}:

\begin{equation}\label{eq:fokker_planck}
\frac{\partial{P}}{\partial{\bar{t}}} + \frac{\partial{S}}{\partial{\bar{F}}}
         =  0, 
\end{equation}
\begin{equation}\label{eq:fokker_planck_flux}
 S(\bar{t},\bar{F}) = \left[- \bar{F}+ R \bar{d}' + 
   \bar{d}-\bar{x^0}- R\frac{\partial}{\partial{\bar{F}}} \right]
   P(\bar{t},\bar{F}). 
\end{equation}
Here, $S(\bar{t},\bar{F})$ is the flux of $P(\bar{t},\bar{F})$. 
To complete the problem, we need boundary conditions at $\bar{F}=0$
and $\bar{F} = -\infty$  $\left(\bar{F} \in (-\infty,0]\text{, no tension at 
contacts}\right)$. At $\bar{F}=-\infty$,
both the probability density and flux
have to vanish. The boundary condition at $\bar{F}= 0$ is nontrivial. 
The flux there depends on the sliders coming into and going out of contact,
since this corresponds to the force changing between 
zero and nonzero values. We know of no way of explicitly deriving this 
boundary condition. 

Let us look at the steady state. At steady state, $ \frac{\partial{P}}{\partial
{\bar{t}}} = 0$ and $\bar{d}' = 0$. From Equation (\ref{eq:fokker_planck}),
the flux is uniform in the domain and $S(-\infty) = 0$ implies it is zero 
everywhere. Setting Equation (\ref{eq:fokker_planck_flux}) to zero, 
the probability density at steady state is: 
\begin{equation}\label{eq:fp_steady_state_pdf}
P(\bar{F}) = c e^{-\frac{\bar{F}^2}{2R}+\frac{(\bar{d}-\bar{x^0})\bar{F}}
{R}}.
\end{equation}
There are two unknowns here, the constant $c$ and the steady state dilatation 
$\bar{d}$. We have one constraint, that the first moment of the probability
density be equal to the applied normal force. Using the constraint,
we can relate the constant $c$ and the dilatation $\bar{d}$ as, 
\begin{equation}\label{eq:fp_steady_state_pdf_constant_determination}
 c = \frac{\bar{F}_N}{\int_{\bar{d}}^\infty \bar{F} e^{-\frac{\bar{F}^2}{2R}+
       \frac{(\bar{d}-\bar{x^0})\bar{F}}{R}} \mathrm{d}\bar{F} }.
\end{equation}
The other constraint comes from the boundary condition at $\bar{F} = 0$ which
cannot be determined explicitly.
 
\subsubsection{Monte Carlo method}
Since the Fokker-Planck equation can be derived only a particular velocity, we
resort to Monte Carlo simulation in the general case. We generate an ensemble 
of sliders, each sliding on an independent 
realization of the noise $\bar{y}(\bar{t})$ with the given statistical 
properties of Gaussian height distribution and exponential correlation. Again, 
we assume the parameters $k_1, k_2, \eta, x^0$ to be the same 
for all the sliders.  The noise is generated at a finite discretization 
size and this introduces a low-wavelength cutoff in the surface features. 
The power spectrum of the Ornstein-Uhlenbeck noise is given by:
$$ S(\omega) = \frac{\lambda \sigma^2}{\pi}\frac{1}{1+\omega^2 \lambda^2},$$
where $\omega$ is the spatial frequency. The power in 
frequencies beyond 10 times the inverse correlation length is small. 
Hence, we discretize the surface to resolve this frequency \cite{ogilvyJA:1}.
For heights of the surface between the discretization points, and the derivative
of the surface height, we use spline interpolation. 

At $\bar{t}=0$, the ensemble of sliders is brought into contact with the rigid 
surface and this results in a global normal force $\bar{F}_N$. In experiments, 
the surfaces slide at a constant global normal force. As in the static contact
case, the dilatation $\bar{d}$ evolves to satisfy the constant normal force 
constraint. At each time step, the rate 
of dilatation is determined to ensure that the global 
normal force remains constant. Once the rate of dilatation is known, the forces and lengths 
of all the sliders can be updated. For time stepping, we use a first order Euler
method. 
Since, at any instant, the state of all the sliders and thus the force
$\bar{F}$ on each of them is known, we can determine the local shear
forces using the local friction law, and, adding them, the global shear force 
$\bar{F}_S$. The friction coefficient is then determined as 
$\mu_k(t) = F_S(t)/F_N$.

\subsubsection{Results}
\paragraph{Test of Monte Carlo method:} 
We use the Fokker-Planck equation of Section \ref{subsec:FokkerPlanck}
as a test of our Monte Carlo method. 
Starting at an initial state, we let the system slide at the velocity 
$\bar{v} = \bar{\lambda}/R$ (the case for which we can derive the Fokker-Planck 
equation) till it reaches steady state. Knowing the normal forces on all the 
sliders, we can compute the probability density. This density is also known from
the solution of the Fokker-Planck equation at steady state 
(Equations (\ref{eq:fp_steady_state_pdf}) and 
(\ref{eq:fp_steady_state_pdf_constant_determination})) and the two can be
compared.

\begin{figure}[h]
\begin{center}
  \includegraphics[scale=0.5]{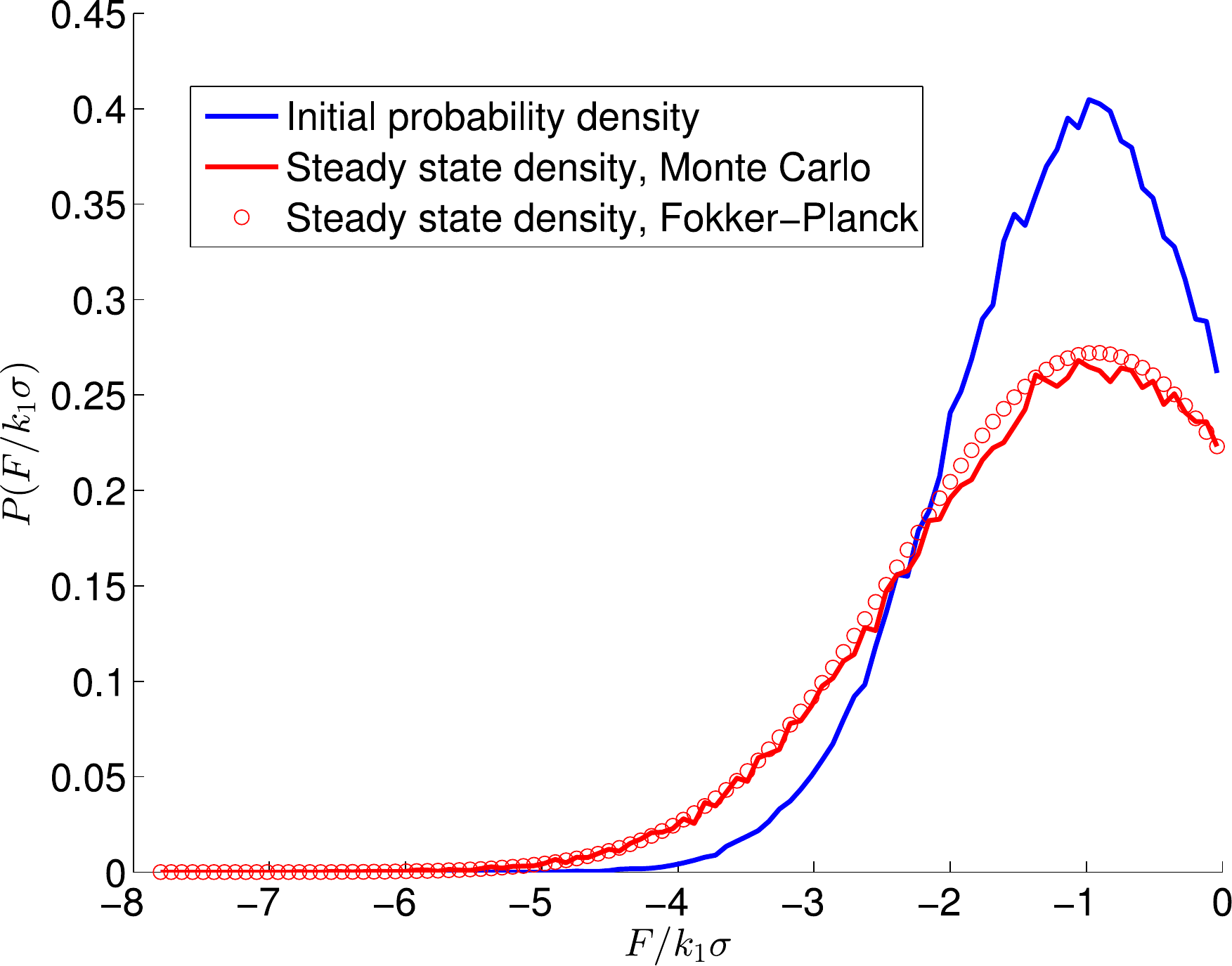}
\caption{Probability density of normal forces at an initial state, and at 
        steady state using two different methods, a Monte Carlo method and 
        the Fokker-Planck solution.}
\label{fig:mc_fp_check}
\end{center}
\end{figure}

Figure \ref{fig:mc_fp_check} shows the probability density of forces in the 
initial and steady states using the Monte Carlo and Fokker-Planck 
methods. In this simulation, $R=2, \bar{\lambda} = 1$ and thus $\bar{v}=0.5$.
An ensemble of $10^5$ sliders is used. From the figure, there is 
a good match in the probability densities using the two methods. As a
further verification, we have also computed the transient evolution with the two 
methods, using the Monte Carlo simulation to prescribe the boundary condition
at $\bar{F}=0$ for the Fokker-Planck equation. The two transients also show a 
good match with each other. 

\paragraph{Velocity jump test:}
Drawing confidence from the above result, we perform  velocity jump simulations
using the Monte Carlo approach. For typical surfaces, $\sigma$ is of the order
of a micron and sliding velocities in the jump tests are usually between
$0.01 \mu$m/s and $100 \mu$m/s. $\eta/k_2$ is of the order of a few seconds for
the two materials mentioned in section \ref{subsuc:singleAsperity}. Thus, 
$\bar{v}$ ranges from $10^{-2}$ to $10^2$.
Starting from an initial state, we let an ensemble of $10^5$ elements
slide at $\bar{v}=0.1$ till it reaches steady state. The velocity is then 
instantaneously changed to $\bar{v} = 1$ and the system is allowed to evolve to 
steady state. Velocity jumps to $\bar{v} = 0.1$ and $\bar{v} = 1$ are then 
repeated.

\begin{figure}[h!]
\centering
\begin{subfigure}[t]{0.4\textwidth}
\includegraphics[width=\textwidth]{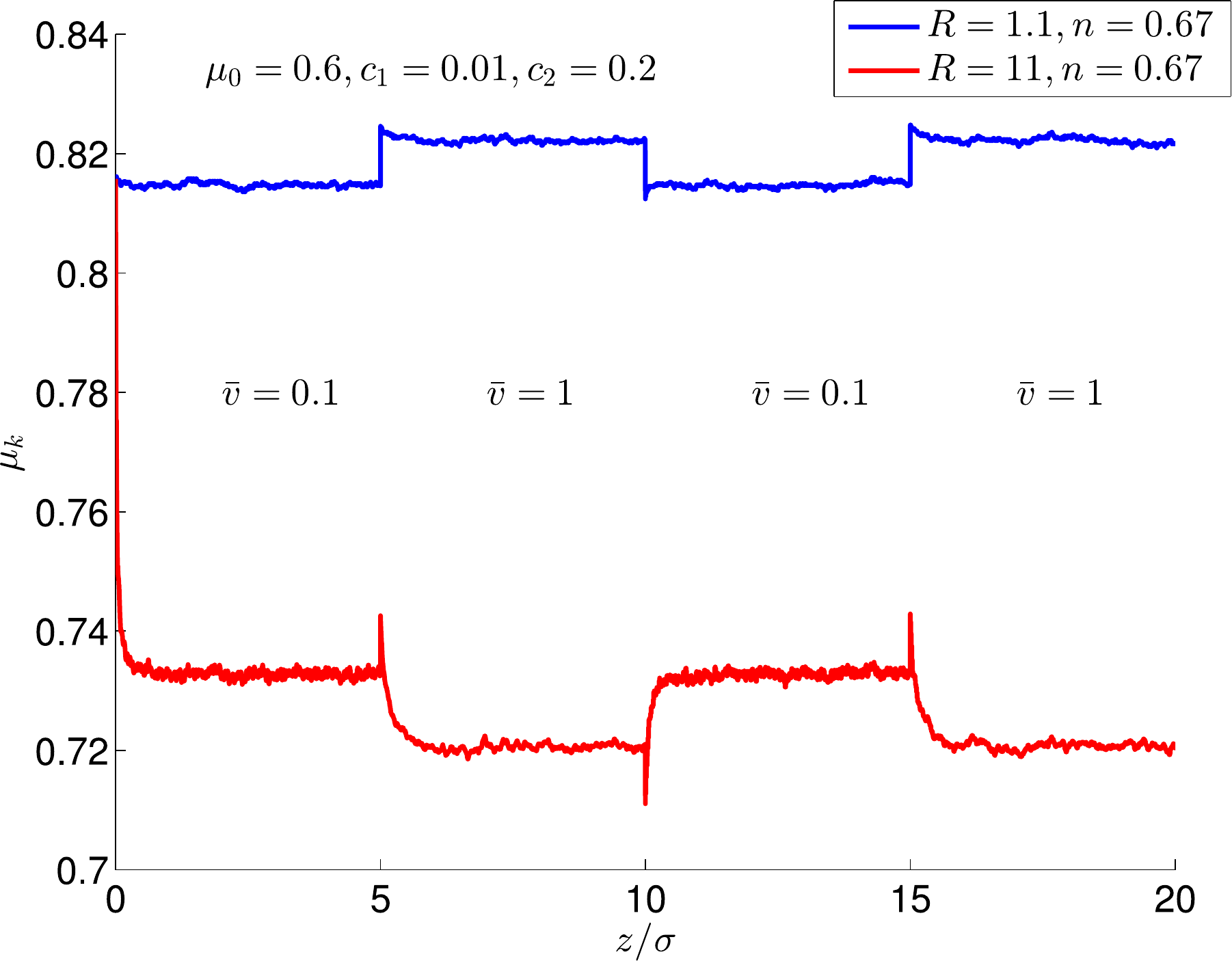}
\subcaption{}
\label{fig:frictionCoefficientEvolutionVelocityJump}
\end{subfigure} 
\quad
\begin{subfigure}[t]{0.4\textwidth}
\includegraphics[width=\textwidth]{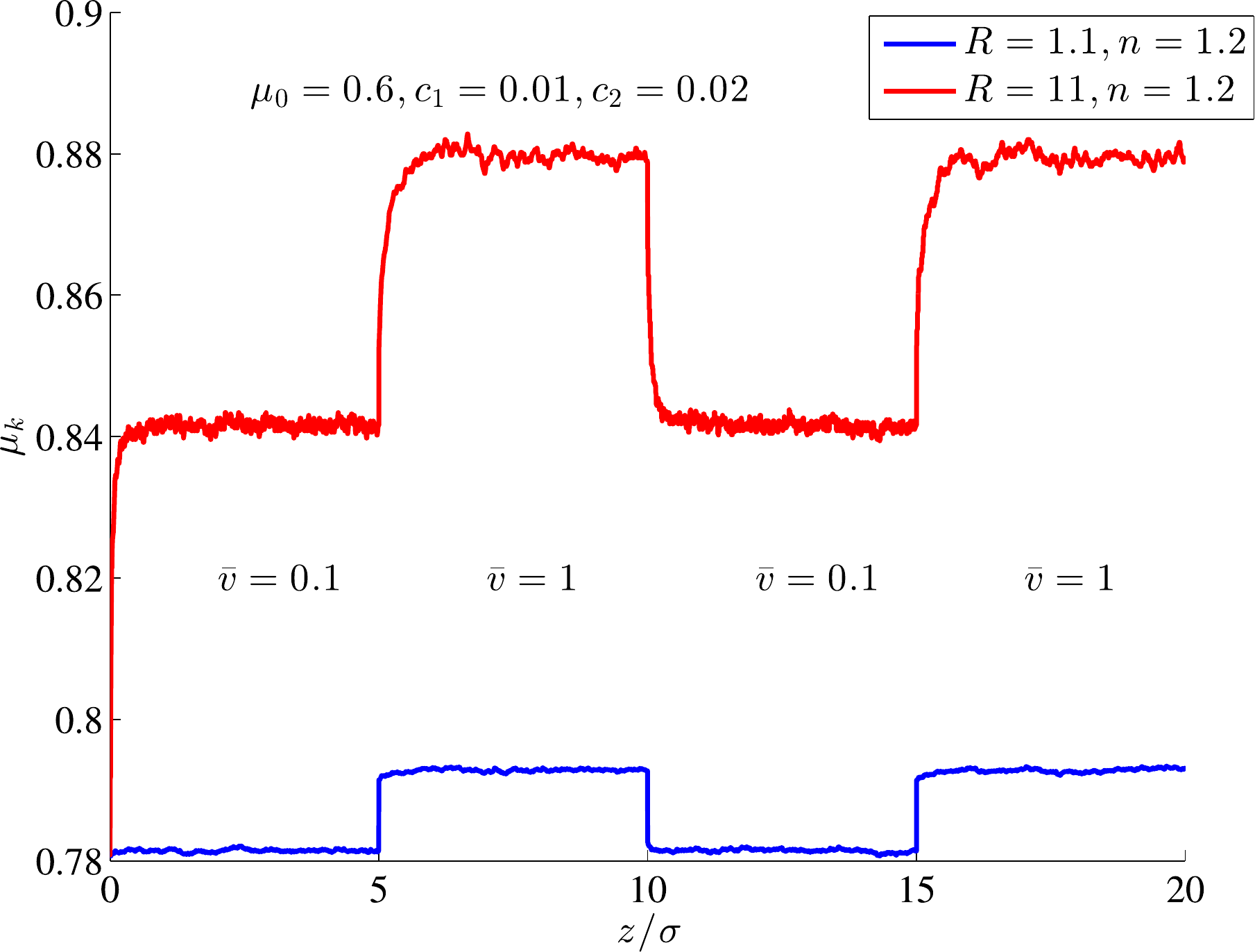}
\subcaption{}
\label{fig:frictionEvolutionPowerGreaterThan1}
\end{subfigure} 
\caption{Evolution of friction coefficient in velocity jump tests. (a) For
         $n<1$, the direct effect and the transient compete against each other.
         Depending on the parameters, either the instantaneous or the transient 
 effect dominates, leading respectively to velocity strengthening or velocity
         weakening. (b) For $n>1$, the friction coefficient changes in the same 
direction during the direct effect and the transient.}
\end{figure}

Figure \ref{fig:frictionCoefficientEvolutionVelocityJump}
shows the evolution of the friction coefficient for two sets of parameter 
values (all parameters are the same except $R$).  In both cases, we find that
the  friction coefficient changes instantaneously when the velocity jumps, and 
the change has the same sense as that of the velocity jump; i.e., the friction
coefficient jumps up (down) when the velocity jumps up (down). 
Following the standard rate and state terminology, we call this jump the 
{\it direct effect}.  This jump is followed by an evolution towards a steady 
state.  We call this the {\it transient}.  
In both cases, this transient changes the friction coefficient in the direction
opposite to the direct effect, i.e., the friction coefficient decreases from
the high value following a jump up and increases from a low value
following a jump down.  
In one case ($R = 1.1$, or the higher curve), the transient is smaller
than the direct effect so that the steady state value is still higher (lower) 
for an increase (decrease) in velocity. 
This represents {\it velocity strengthening} behavior.  In the other case 
($R = 11$, or the higher curve), the transient is larger than the direct effect
so that the steady state value is lower (higher) for an increase (decrease)
in velocity. This represents {\it velocity weakening} behavior.
If the power $n$ in the local friction law is greater than $1$, 
as shown in Figure
\ref{fig:frictionEvolutionPowerGreaterThan1}, the friction
coefficient change during the transient is in the same direction as the 
direct effect.   

\begin{figure}[h!]
\begin{center}
  \includegraphics[scale=0.5]{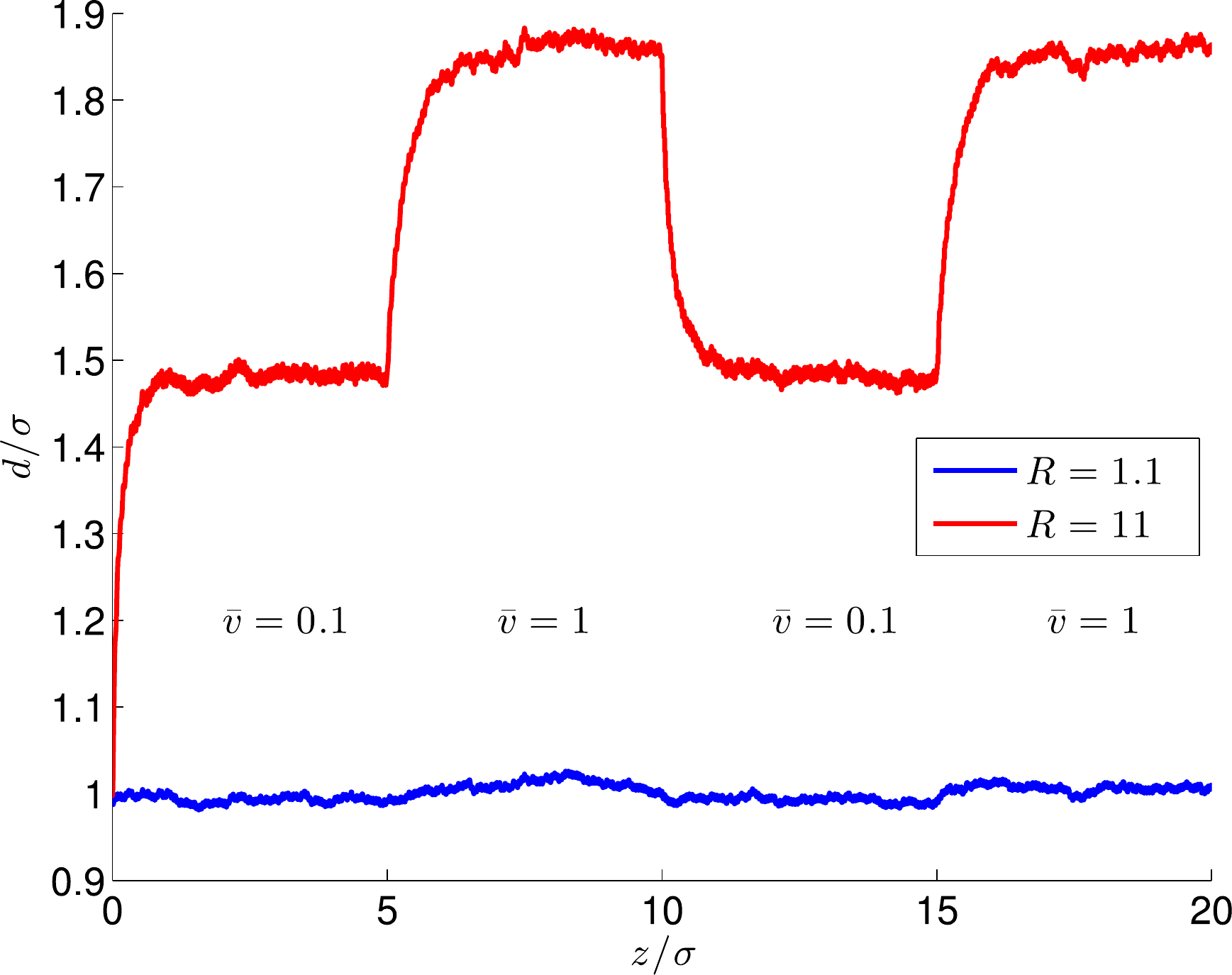}
\caption{Evolution of dilatation in the velocity jump test. The average normal force on 
         an asperity in contact is higher at higher sliding speeds. Thus, for the same
         global normal force, fewer sliders are in contact at higher speeds and the 
         dilatation is larger. For $R$ close to $1$, the changes in dilatation
are hardly apparent.}
\label{fig:dilatationEvolutionVelocityJump}
\end{center}
\end{figure}

Figure \ref{fig:dilatationEvolutionVelocityJump}
shows the evolution of the dilatation for the same two sets of parameter
values.  We observe that the dilatation changes continuously with no 
jumps, and evolves towards a steady state following an imposed 
velocity jump.  Further, the evolution occurs in the same sense, toward higher
dilatation for higher velocity and vice versa, for both sets of parameters.  
The excursions are larger for larger $R$ (the change in dilatation is 
hardly visible for the case $R = 1.1$).

In our model, the force on an asperity changes continuously with time, and 
therefore the force distribution also changes continuously with time.  This is
reflected in Figure \ref{fig:dilatationEvolutionVelocityJump}.
Therefore, the direct effect in Figure 
\ref{fig:frictionCoefficientEvolutionVelocityJump} is a direct
consequence of our local friction law.  In fact, the instantaneous increase in 
$\mu_k$ is given by $c_1 \log (v_{new}/v_{old})$. 
The subsequent evolution is a result of
the collective behavior due to the evolution of the force distribution in 
addition to the local friction law.  Since the dilatation increases with 
increasing velocity, the applied global normal force is carried by fewer
asperities with larger average forces on each 
(we elaborate on this later).  
Consequently, if the exponent $n$ in the local friction law satisfies $n<1$ as 
we expect from the physics, velocity jump will lead to a decrease in friction 
coefficient during the evolution phase and velocity jump down will lead to an 
increase in friction coefficient during evolution phase.  In short, the direct
effect and evolution will always compete with each other.

The above results capture many features of experimental observations.
Both the direct effect and the transient are observed in experiments 
\cite{dieterich:2,dieterich:3,marone:1}. Further, the direct effect always
follows the velocity jump. Furthermore, the direct effect and evolution always
change the friction coefficient in opposite directions. This is consistent with
the requirement that $n \in [2/3,1]$.  Finally, both velocity-
strengthening and velocity-weakening behaviors have been observed.

\paragraph{Velocity strengthening vs. velocity weakening:}

\begin{figure}[h!]
\centering
\begin{subfigure}[t]{0.4\textwidth}
\includegraphics[width=\textwidth]{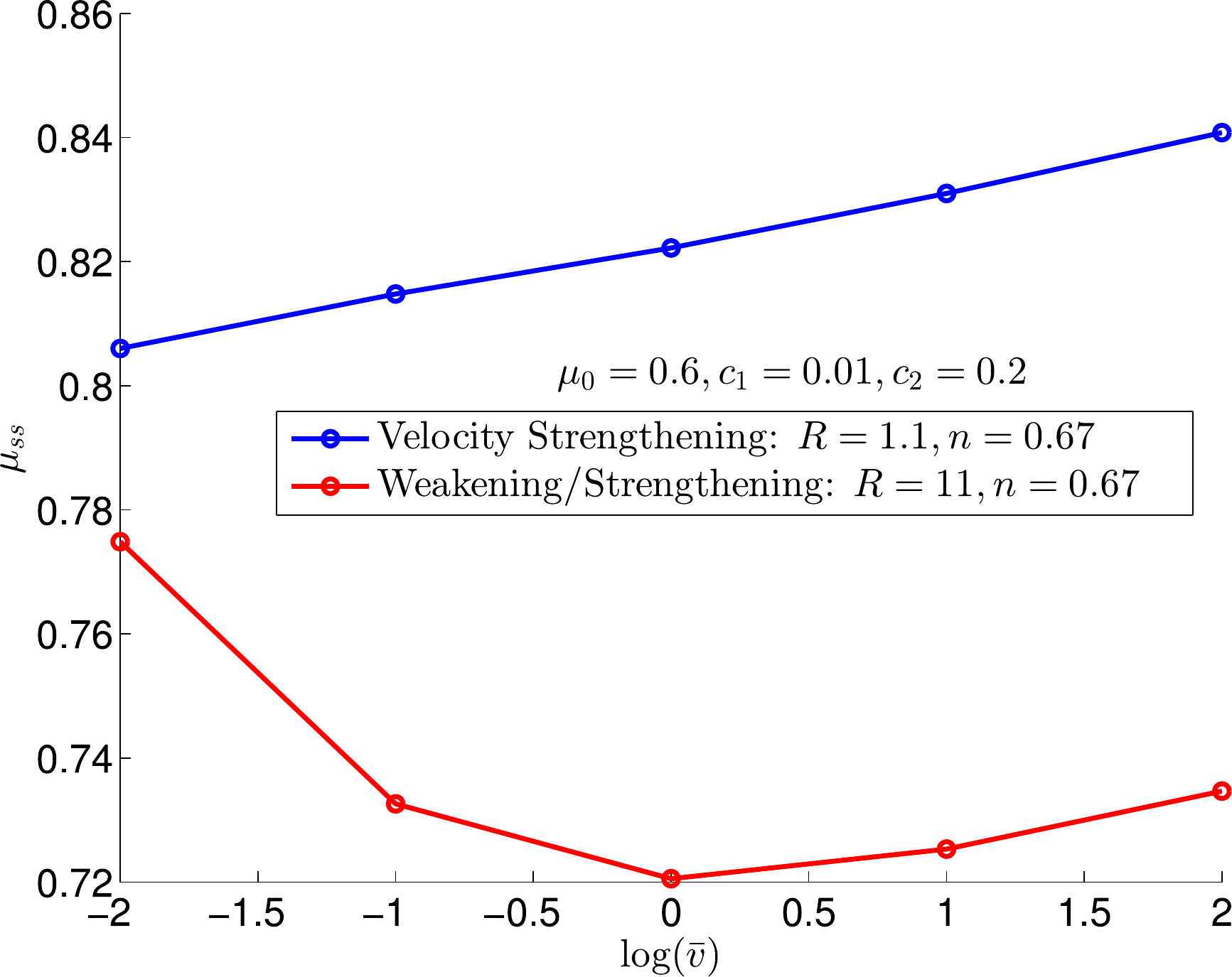}
\subcaption{}
\label{fig:frictionCoefficientSteadyStateVelocityDependence}
\end{subfigure} 
\qquad
\begin{subfigure}[t]{0.4\textwidth}
\includegraphics[width=\textwidth]{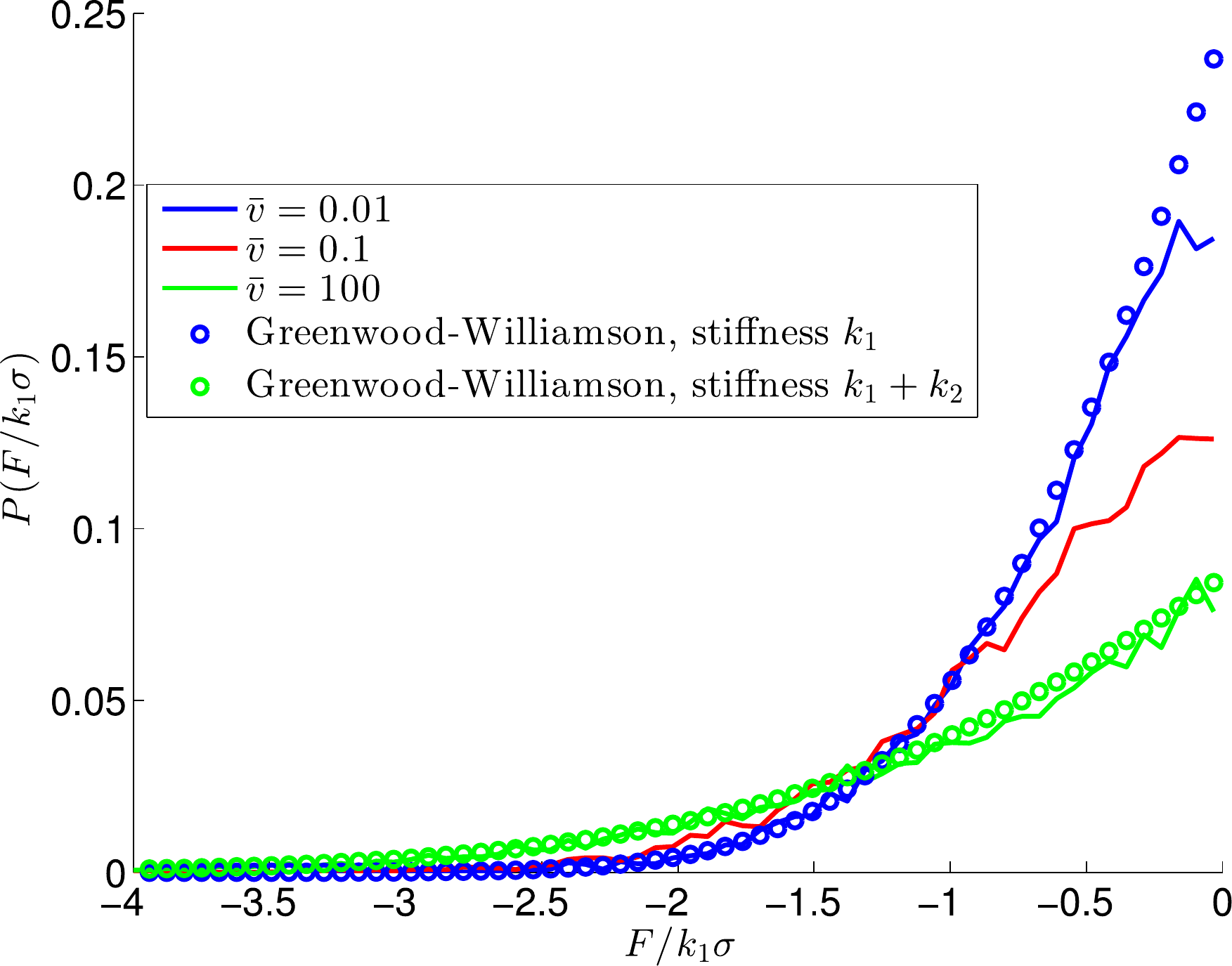}
\subcaption{}
\label{fig:probabilityDensityForceSteadyStateVelocityDependence}
\end{subfigure} 
\caption{(a) Velocity dependence of the steady-state friction coefficient.
Depending on the parameters and the sliding speed, we can have velocity 
strengthening or velocity weakening. Transitions between strengthening and 
weakening behaviors have been observed in experiments \cite{shimamoto:1}
(b)Probability density of the normal forces at steady state at different
sliding speeds (solid lines). At higher speeds, the average normal force on an 
asperity in contact is higher. Thus, the area under the curve, which represents 
the fraction of sliders in contact, is smaller at higher speeds. The first
moment of the force distribution, which is the global normal force, is the same 
for all the curves. Also shown is the probability density of
         forces in a linearized Greenwood-Williamson model with stiffnesses
         $k_1$ and $k_1+k_2$. These are the force distributions in the limit
of $0$ and infinite sliding speeds.}
\end{figure} 

Figure \ref{fig:frictionCoefficientSteadyStateVelocityDependence} 
shows the dependence of the steady-state friction coefficient on the sliding 
speed. One case (the blue curve) shows velocity strengthening at all sliding 
speeds whereas the other case (the red curve) shows velocity weakening at low
sliding speeds and strengthening at higher speeds. Since the SLS has only one
relaxation timescale, the distribution of forces on asperities is sensitive to
the sliding velocity only in a limited range of velocities (the lower 
velocities for the red curve). Outside this velocity range (the higher 
velocities for the red curve), the local friction properties dominate and we get
velocity-strengthening behavior. The transition between velocity strengthening 
and velocity weakening has been observed in experiments \cite{shimamoto:1}.

\paragraph{Distribution of forces on asperities:}
Figure \ref{fig:probabilityDensityForceSteadyStateVelocityDependence} shows the probability
density of the normal forces at steady state at different sliding speeds. At 
higher sliding speeds, for the same global normal force, fewer sliders need to
be in contact since the normal force on each of the ones in contact is 
higher on average. 
This can be seen in the figure where the area under the probability density 
curve, which represents the fraction of sliders in contact, is smaller at 
higher speeds.
Also shown is the probability density of the normal forces in a linearized 
version of the Greenwood-Williamson model (where the nonlinear Hertzian 
contact is replaced by a linear one). 
The two curves correspond to two different stiffnesses, $k_1$
and $k_1+k_2$. At low sliding speeds, the SLS elements have more
time to evolve towards their steady state and thus effectively only the spring
with stiffness $k_1$ is active during contact. At high speeds, the dashpot has 
little time to react, and the effective
stiffness is nearly $k_1+k_2$. This can be seen in Figure 
\ref{fig:probabilityDensityForceSteadyStateVelocityDependence} where the probability 
densities at low and high sliding speeds are similar to the densities
of the GW model with stiffnesses $k_1$ and $k_1+k_2$, respectively.
The probability densities at different speeds can be mapped to 
the probability density of the Greenwood-Williamson model with the effective 
stiffness dependent on the sliding speed.

\begin{figure}[h!]
\centering
\begin{subfigure}[t]{0.4\textwidth}
\includegraphics[width=\textwidth]{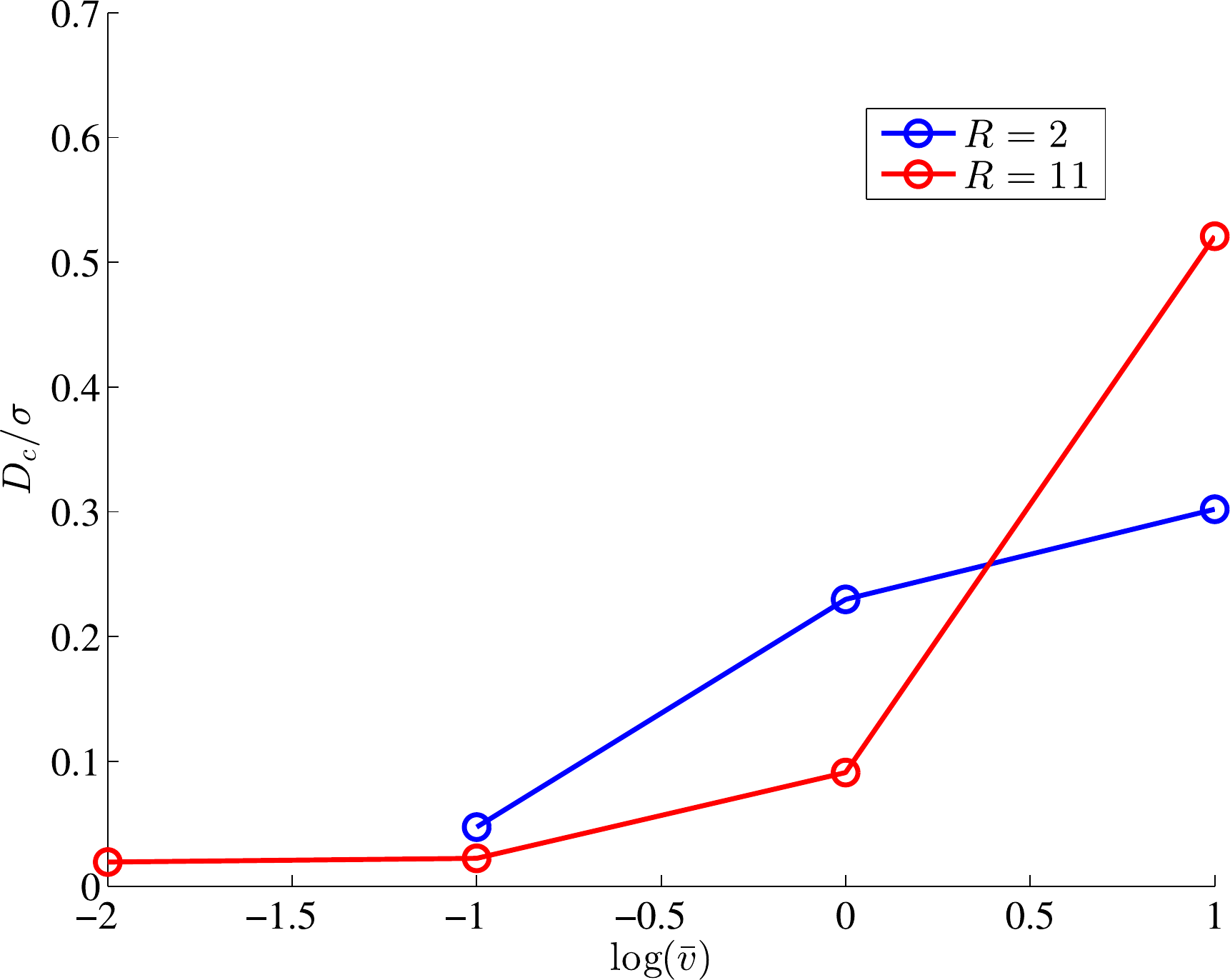}
\subcaption{}
\label{fig:decayLength}
\end{subfigure} 
\qquad
\begin{subfigure}[t]{0.4\textwidth}
\includegraphics[width=\textwidth]{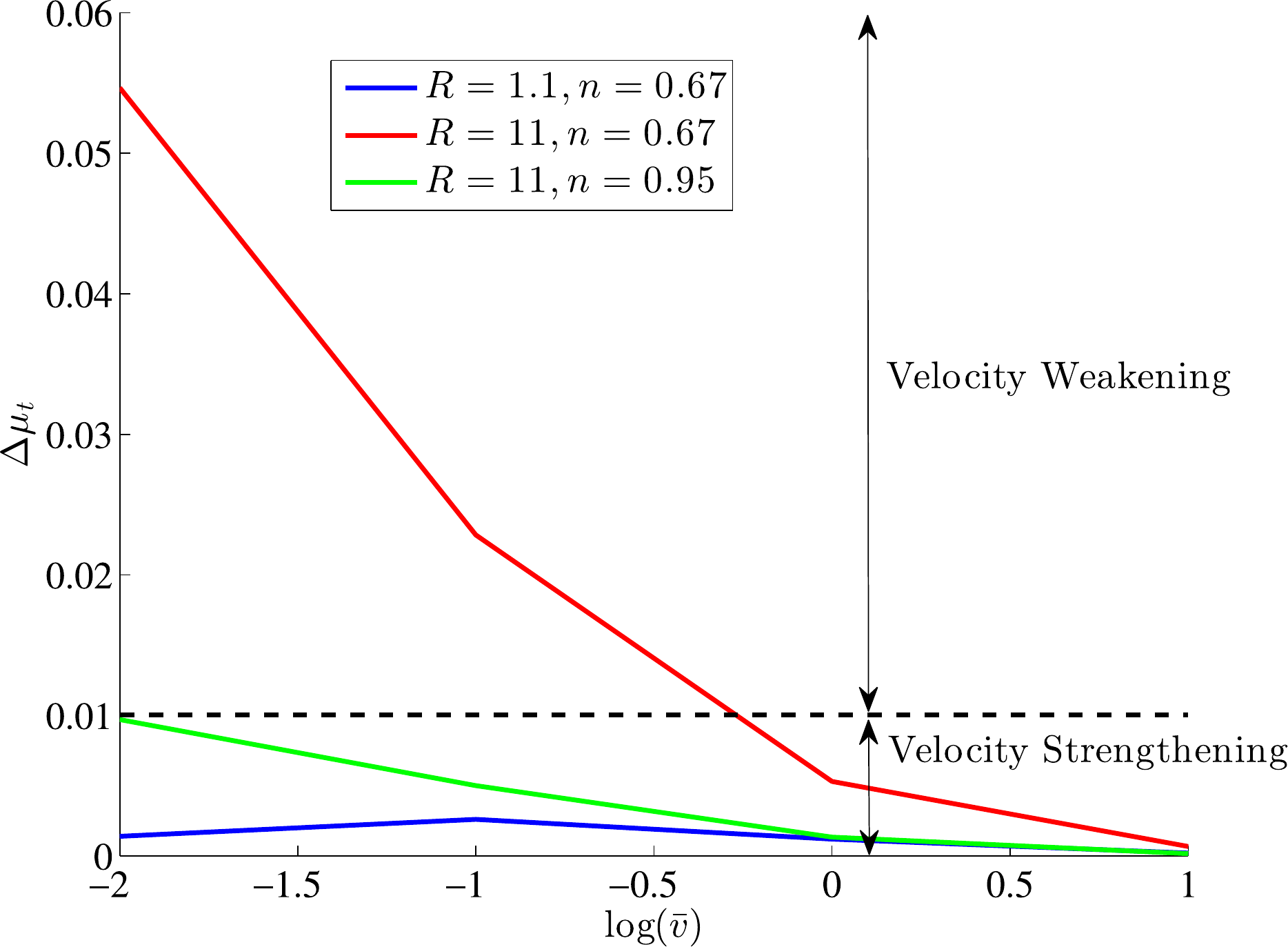}
\subcaption{}
\label{fig:deltaMuTransient}
\end{subfigure} 
\caption{(a)The characteristic length over which the system decays to
         steady state  depends on the stiffness ratio $R$ and the sliding speed
         $\bar{v}$. In many experiments, this distance is found to be a constant that 
         depends only on the roughness of the surfaces.
         (b)The change in the friction coefficient during the transient phase
        following the direct effect, for three different combinations of $R$ and $n$.
        If the change during the transient is greater than the direct effect, we have
        velocity-strengthening (above the dashed line); otherwise, we have velocity-
        weakening.}
\end{figure}

\paragraph{Characteristic slip distance:}
$D_c$ is the characteristic length 
scale over which the system evolves to its steady state in velocity jump tests.
We calculate this by fitting an exponential to the evolution of
$\mu_k$. 
Figure \ref{fig:decayLength} shows the dependence of
$D_c$ on the sliding speed $\bar{v}$ for two different values of $R$ and for
$n = 0.67$ and $\bar{\lambda} = 1$. 
$D_c$ also depends on the correlation length of the surface,
$\bar{\lambda}$. Recall that the parameter $\bar{\lambda}$ appears in the equations only
as $\bar{v}/\bar{\lambda}$. If we have two surfaces with correlation
lengths $\bar{\lambda}_1, \bar{\lambda}_2$, and the SLS ensemble slides 
on these surfaces at velocities $\bar{v}_1$ and $\bar{v}_2$ such that 
$\bar{v}_1/\bar{\lambda}_1=\bar{v}_2/\bar{\lambda}_2$, then the decay
lengths are related as: 
$$D_c(\bar{v}_1,\bar{\lambda}_1) = 
\frac{\bar{\lambda}_1}{\bar{\lambda}_2}D_c(\bar{v}_2,\bar{\lambda}_2) .$$
$D_c$ is fairly independent of $\bar{v}$ in some laboratory experiments 
\cite{dieterich:2}, and our simplified model does not capture this independence.

\paragraph{Parametric study:}
After nondimensionalization, we have five parameters: $R,\bar{\lambda},
n, c_1$ and $c_2$. Let us study the effect of each of them. 

$R = 1+k_2/k_1$ is the ratio between the instantaneous 
and steady-state stiffnesses of the SLS. When $R$ is 
close to $1$, the instantaneous and steady-state responses of the SLS are 
close, the differences between the distribution of forces on asperities
at different sliding speeds is small, and thus, the transient change in $\mu_k$
following a jump in sliding speed is small. By a similar consideration, for large
$R$ $(k_2 \gg k_1)$, the transient change in $\mu_k$ is large. 
Hence, for the same instantaneous effect, as the value of $R$ increases, the 
behavior will change from velocity-strengthening to velocity-weakening, as 
illustrated in Figure \ref{fig:frictionCoefficientEvolutionVelocityJump}.

In the governing equations (\ref{eq:force_langevin_nondim}), (\ref{eq:noise_statistics_nondim}),
(\ref{eq:sls_relaxation_nondim}), the parameter $\bar{\lambda}$ appears only as 
$\bar{v}/\bar{\lambda}$. Thus, we can think of $\bar{\lambda}$ as setting a scale for
the sliding speed. Consider two surfaces with correlation lengths $\bar{\lambda}_1$ 
and $\bar{\lambda}_2$ on
which an SLS ensemble slides at two speeds $\bar{v}_1$ and $\bar{v}_2$. If 
$\bar{v}_1/\bar{\lambda}_1 = \bar{v}_2/\bar{\lambda}_2$, and if we start from the same 
initial state, the evolution of the two systems will be exactly the same. Since we 
have a velocity dependence in the local friction law, the coefficients of friction 
of the two systems will differ by $c_1\log(\bar{v}_1/\bar{v}_2)$.

The power $n$ in the local friction law represents the elasticity/plasticity of 
contacts. When $n=1$, all contacts in the ensemble are plastic and the distribution
of normal forces among the asperities has no effect on the friction coefficient. 
This is equivalent to the Bowden and Tabor model, modulo the velocity-dependent
term in the local friction law. When $n=2/3$, the effect of 
distribution of forces among asperities plays an important role in determining
$\mu_k$.  Parameters $c_1$ and $c_2$ describe the microscopic frictional 
response of a single contact. 

We would like to study how each of the above parameters affects the friction
coefficient. 
To this end, we do velocity jump experiments at different speeds for different 
values of the parameters $R$ and $n$, and calculate the change in $\mu_k$ 
during the transient phase ($\Delta\mu_t$) following the instantaneous jump.  
Figure \ref{fig:deltaMuTransient} shows the value of $\Delta\mu_t$ at different
velocities for three different combinations of parameters $R$ and $n$. 
A value of $c_2 = 0.2$ has been used ($\Delta\mu_t$ changes linearly with $c_2$). 
Also shown as a dashed line is $c_1$ which
has been assumed to be $0.01$. We have velocity strengthening when 
$\Delta\mu_t < c_1$ (below the dashed line) and velocity weakening when 
$\Delta\mu_t > c_1$ (above the dashed line). In many experiments, $\Delta\mu_t$
is observed to be independent of the sliding speed.  
In our model, $\Delta\mu_t$ is not a constant for given material properties 
but varies with the sliding velocity. 

\subsubsection{Comparison with rate and state formulations}
\begin{figure}[h!]
\centering
\begin{subfigure}[t]{0.4\textwidth}
\includegraphics[width=\textwidth]{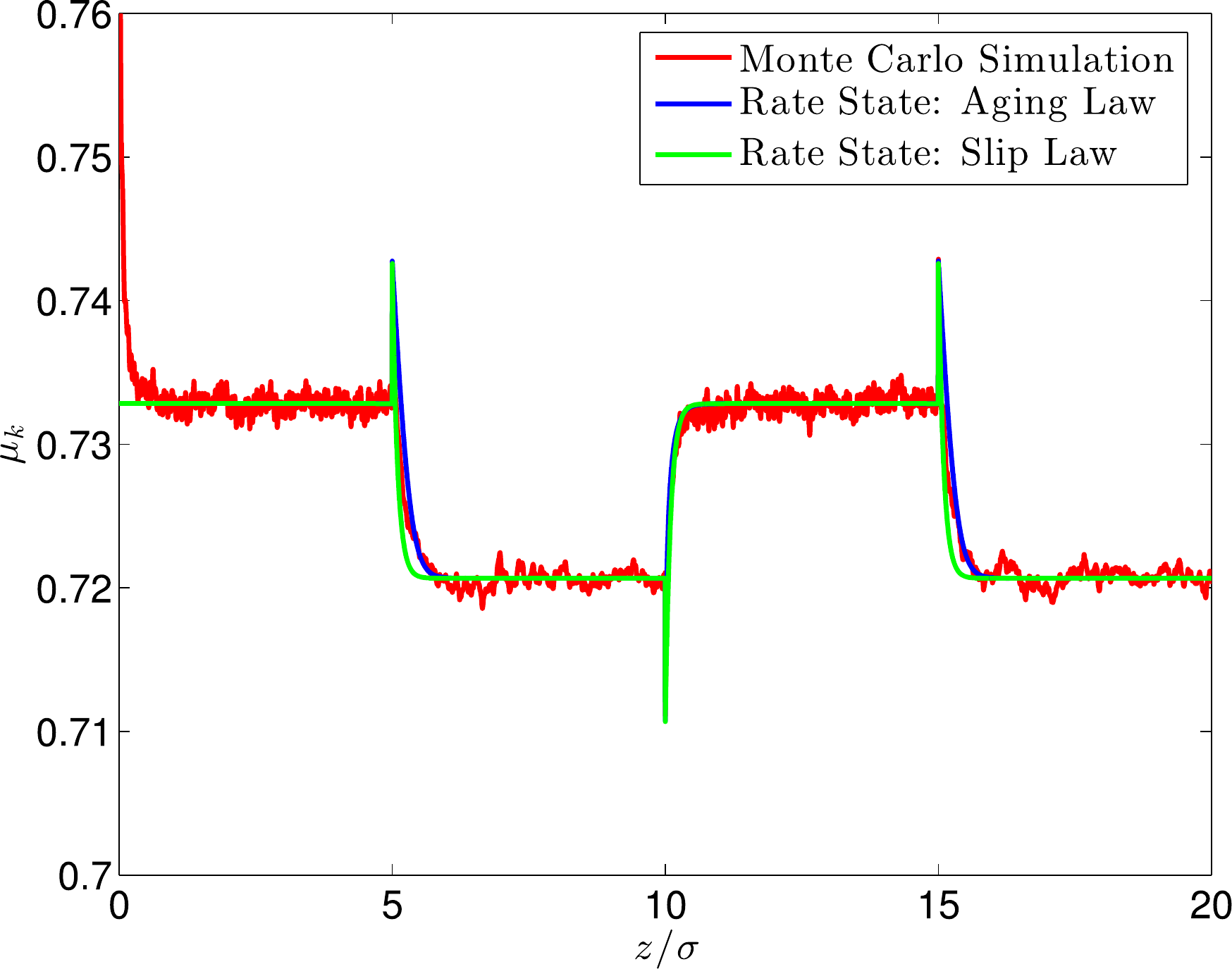}
\subcaption{}
\label{fig:frictionVelocityJumpRateStateFit}
\end{subfigure} 
\qquad
\begin{subfigure}[t]{0.4\textwidth}
\includegraphics[width=\textwidth]{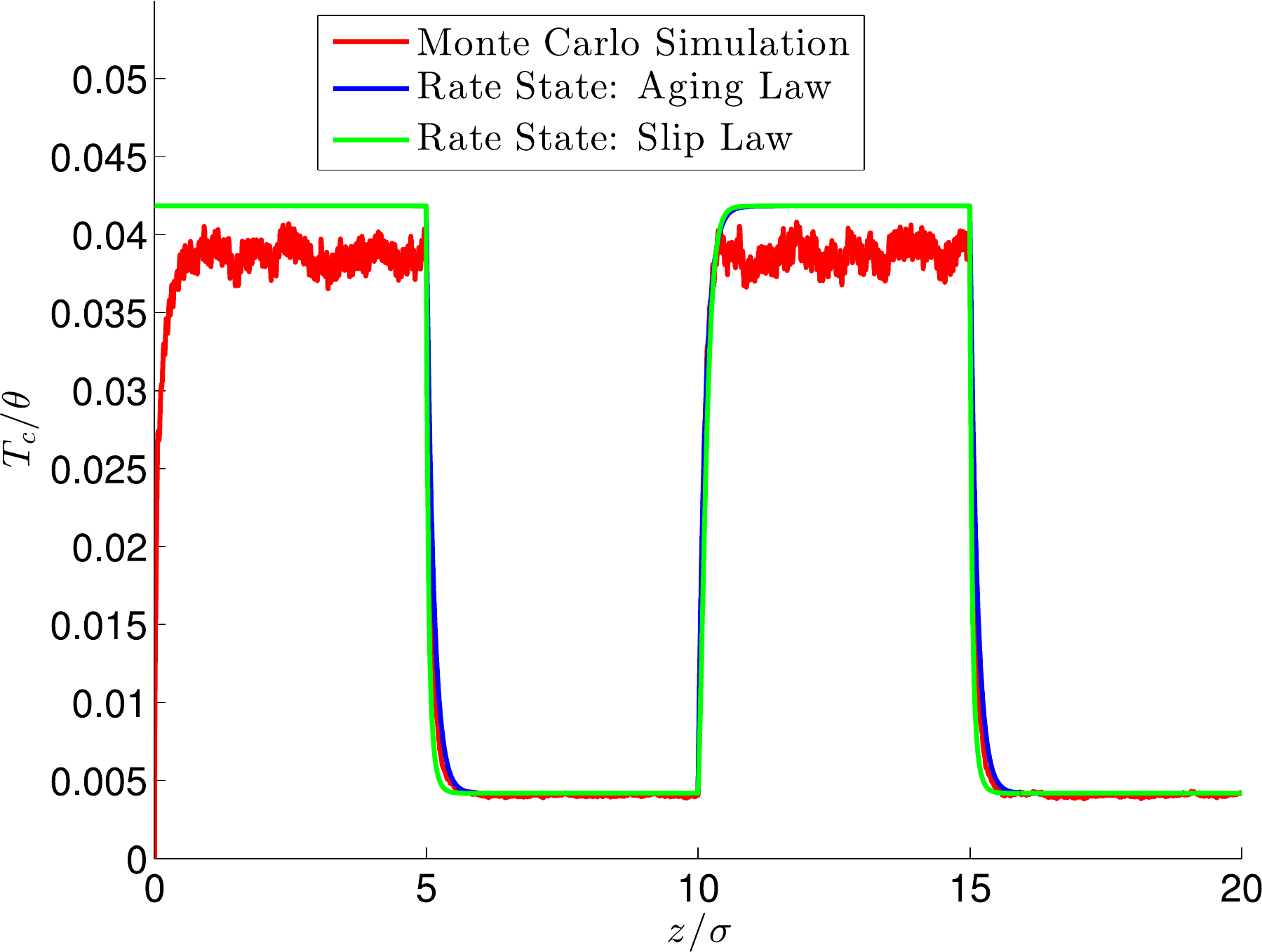}
\subcaption{}
\label{fig:stateVariableAverageTimeContact}
\end{subfigure} 
\caption{(a)Evolution of $\mu_k$ in a velocity jump test (with $R = 11$, 
$n = 0.67$), along with fits for two versions of the rate and state laws. 
(b)The average time of contact calculated using the Monte Carlo simulation
(red), and the evolution of the state variable $\theta$ calculated from the
fits in (a). The average time of contact is indeed close to the state variable
evolution, supporting the interpretation of the state variable in the rate
and state laws as the average contact time.}
\end{figure}

Our results on the evolution of friction coefficient in jump tests can be well
fitted by the rate and state equations (\ref{eq:RS}-\ref{eq:slip_law}), 
as illustrated in Figure \ref{fig:frictionVelocityJumpRateStateFit}. 
In the example shown, the best fit parameters for the 
rate and state equations are: $\mu_0 = 0.7207, a = 0.0043 ( =c_1/\ln10),
b = 0.0096$. $v^*$ was chosen to be $1$ and $D_c/\sigma$ is $0.13$ for the aging
law and $0.042$ for the slip law.

\paragraph{State variable:}
In the rate and state laws, the state variable is sometimes interpreted as the
average contact time of the asperities \cite{dieterich:4,baumberger:2}.
This time of contact can be calculated explicitly in our simulations.  
Figure \ref{fig:stateVariableAverageTimeContact} shows the evolution of the
average time of contact $(T_c)$ in a velocity jump test. Also shown in the 
figure is the evolution of the state variable $\theta$ for the two rate and 
state laws, calculated using the friction evolution
fit of Figure \ref{fig:frictionVelocityJumpRateStateFit} ($\theta$ does not 
match $T_c$ in absolute value, the figure shows $\theta$ scaled to match $T_c$
for the last data point). The steady-state $T_c$ is approximately inversely 
proportional to the sliding speed, as proposed for the state variable in the rate
and state formulations.  Thus, our model is consistent with the idea that state 
variable in the rate and state laws is related to the average contact time.
\begin{figure}[h!]
\centering
\includegraphics[scale=0.5]{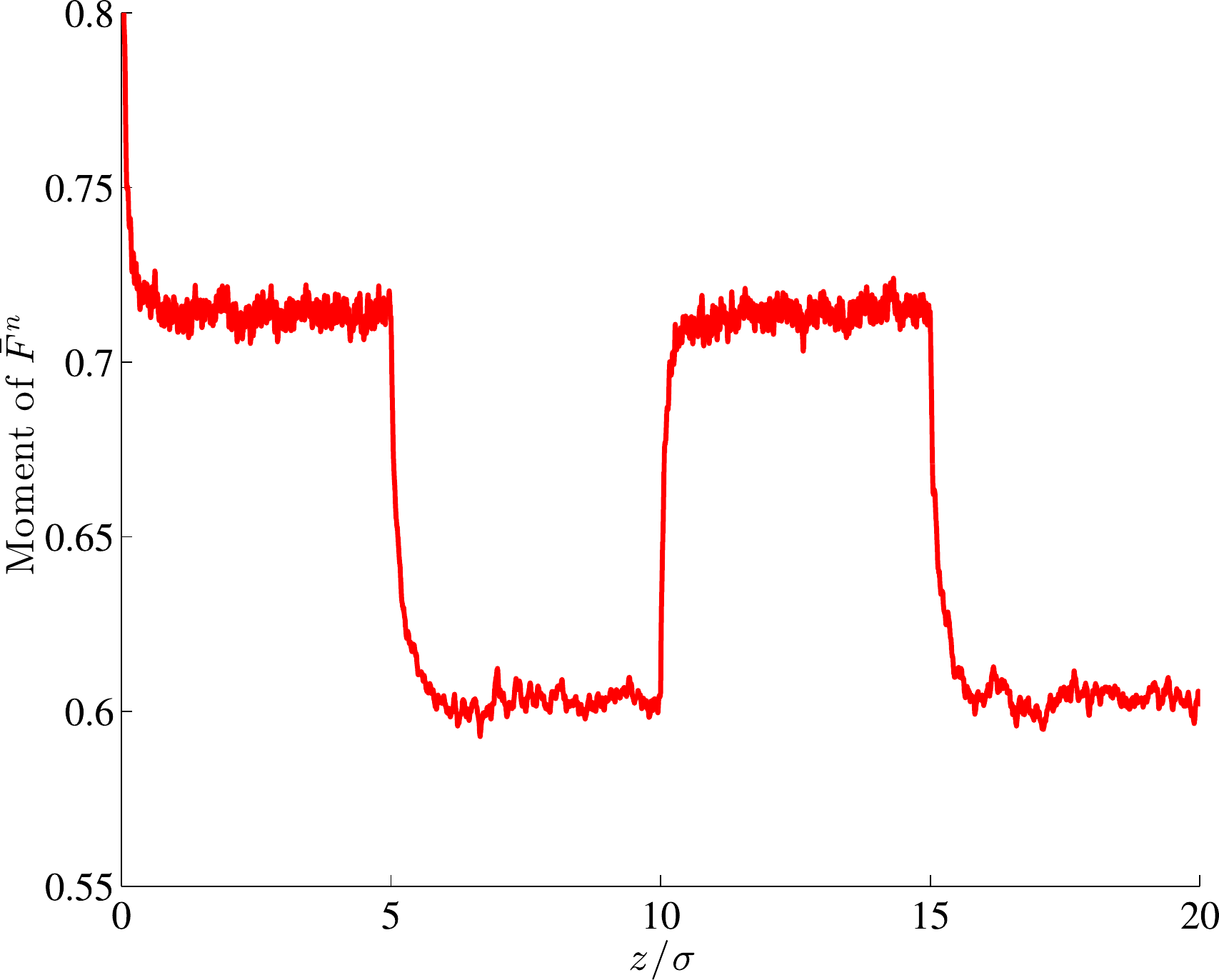}
\caption{Evolution of the moment of $\bar{F}^n$ during the velocity jump test. 
 This moment serves as the state variable in our model.}
\label{fig:momentEvolutionVelocityJump}
\end{figure}
\paragraph{Moment as a state variable:}
In our model, the transient behavior in the velocity jump tests is proportional
to the moment of $\bar{F}^n$, i.e., to
\begin{equation} 
\psi(\bar{t}) = \int_{\bar{F}} \bar{F}^{n}P(\bar{t},\bar{F})\mathrm{d}\bar{F}.
\end{equation} 
The evolution of this moment is shown in Figure 
\ref{fig:momentEvolutionVelocityJump} for $n = 2/3$.
Hence this moment acts as the state variable in our model, 
and we can formulate the following rate and state description of  our model:
\begin{equation} 
\mu_k(\bar{t}) = \mu_0 + c_1 \log(\bar{v}) + c_2 \psi(\bar{t}),
\end{equation}
\begin{equation} 
\psi'(\bar{t}) = g(\bar{v}) (\psi(\bar{t}) - \psi_{ss}(\bar{v})),
\end{equation}
where $g(\bar{v})$ and $\psi_{ss}(\bar{v})$ can be determined
by Monte Carlo simulations for given model parameters $R, n, \bar{\lambda}$.
In fact, $g = \bar{v}/\bar{D}_c$ where $\bar{D}_c$ is shown in Figure 
\ref{fig:decayLength} for $n = 0.67$ for two different values of $R$.

\subsection{Nonlinear contact model}
\begin{figure}[h!]
\begin{center}
  \includegraphics[scale=0.5]{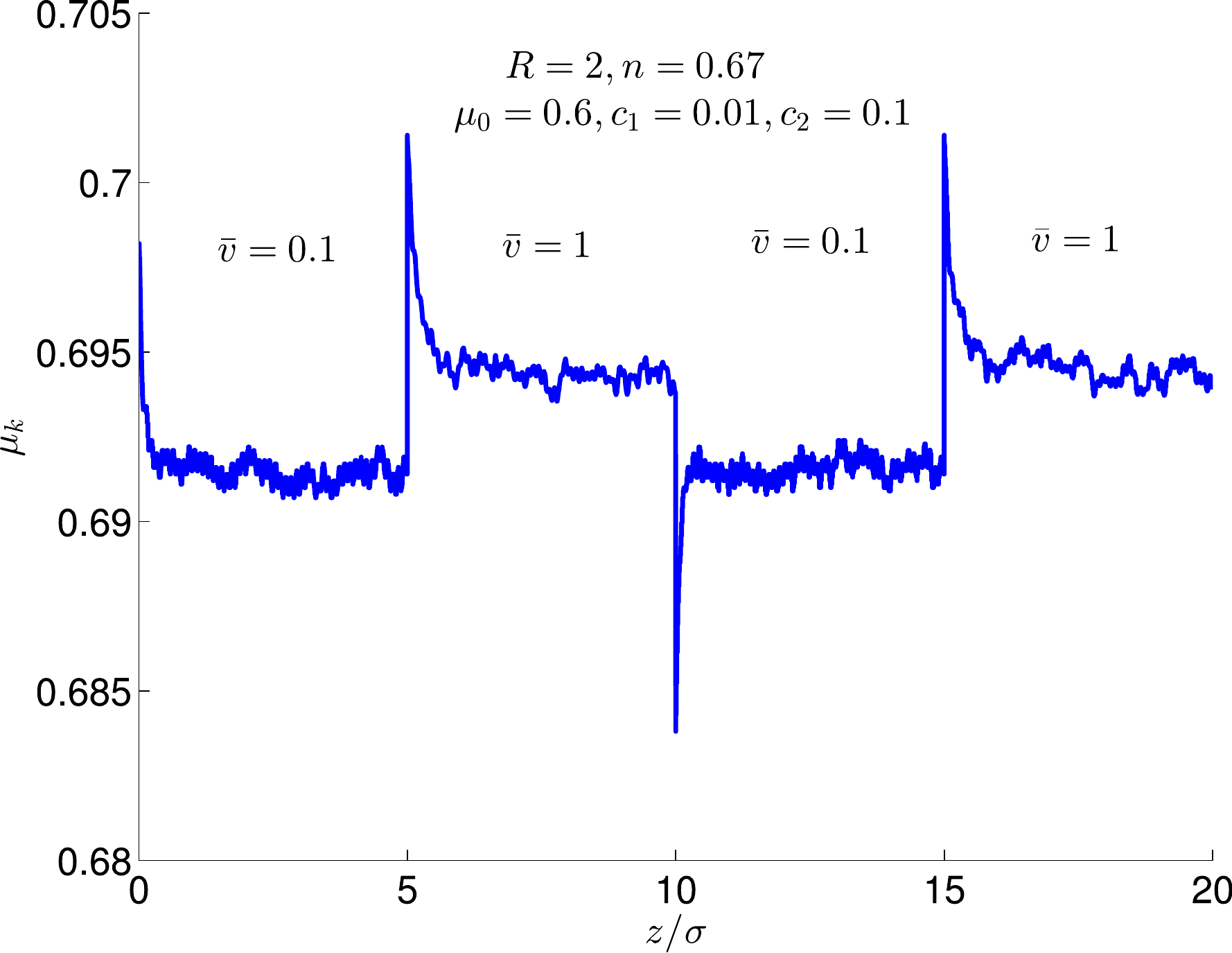}
\caption{Velocity jump experiment with a nonlinear asperity model. The evolution of 
         friction coefficient is qualitatively similar to the evolution in 
 Figure \ref{fig:frictionCoefficientEvolutionVelocityJump}. Though not
conclusive, this suggests that the qualitative features of the results are 
robust with respect to the single asperity model.}
\label{fig:friction_evolution_nonlinear_contact_model}
\end{center}
\end{figure}

Until now, we have modeled asperities as linear viscoelastic elements.  
To test how sensitive the results of our model are to this assumption, 
we repeat the velocity jump simulations with a modified model for an asperity.
We make spring $1$ in the SLS nonlinear. The constitutive equations are:
$$ F = k_1 sgn(x-x^0) |x-x^0|^{3/2} + \eta \dot{x}_\eta, $$ 
$$ \eta \dot{x}_\eta  = k_2(x_2-x_2^0). $$ 
The power $3/2$ has been chosen to  mimic Hertzian contact behavior. 
Using the Monte Carlo method, we repeat the velocity jump experiment. 
Figure \ref{fig:friction_evolution_nonlinear_contact_model} shows that the 
evolution of the friction coefficient is similar to the linear SLS case. 
While not conclusive, this example suggests that the 
qualitative features of the results are not crucially dependent on the 
particular description of a single asperity.

\subsection{Multiple timescales}\label{subsec:multipleTimescales}
\begin{figure}[h!]
\centering
\begin{subfigure}[t]{0.4\textwidth}
\includegraphics[width=\textwidth]{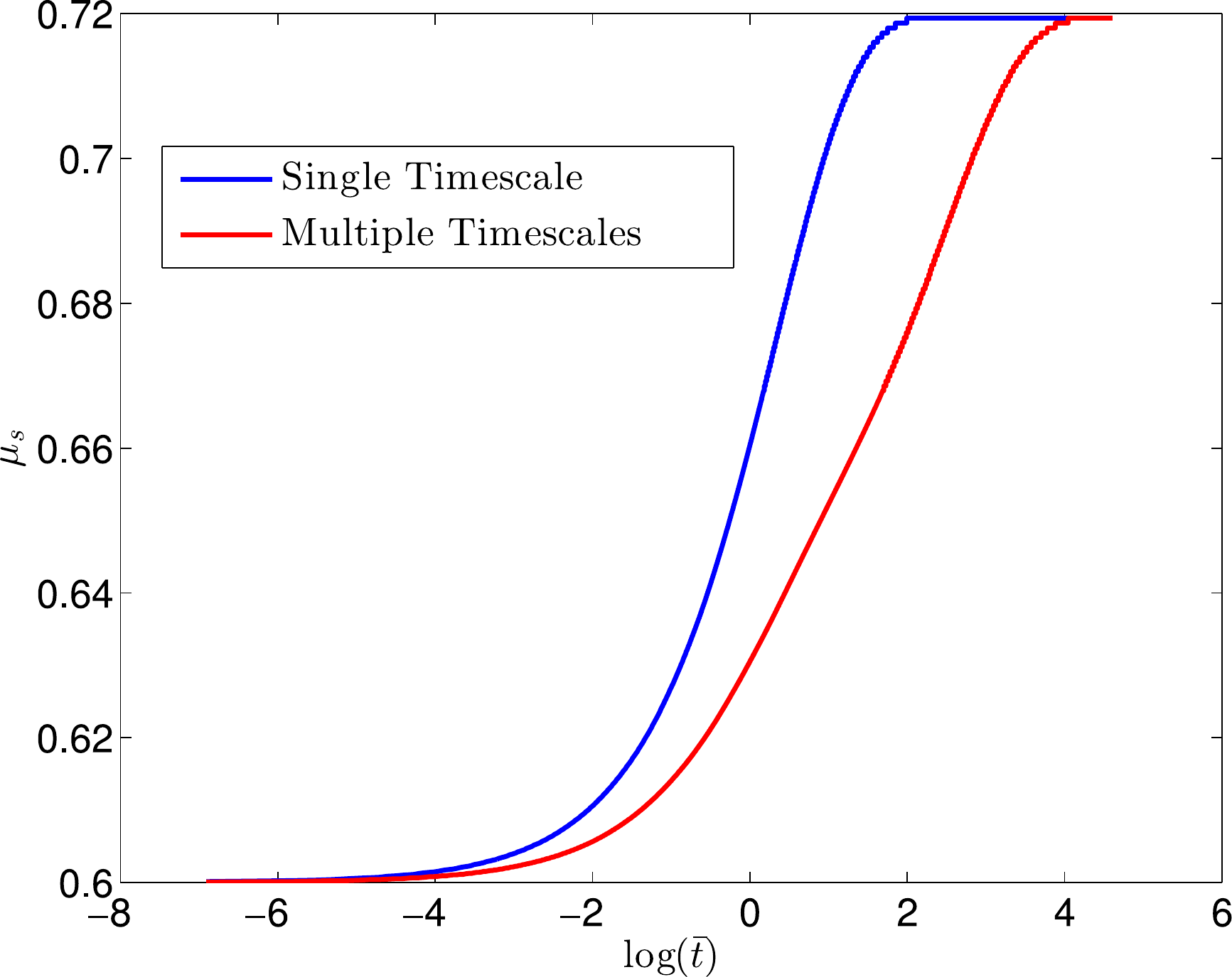}
\subcaption{}
\label{fig:frictionEvolutionStaticContactSingleMultipleTimescales}
\end{subfigure} 
\qquad
\begin{subfigure}[t]{0.4\textwidth}
\includegraphics[width=\textwidth]{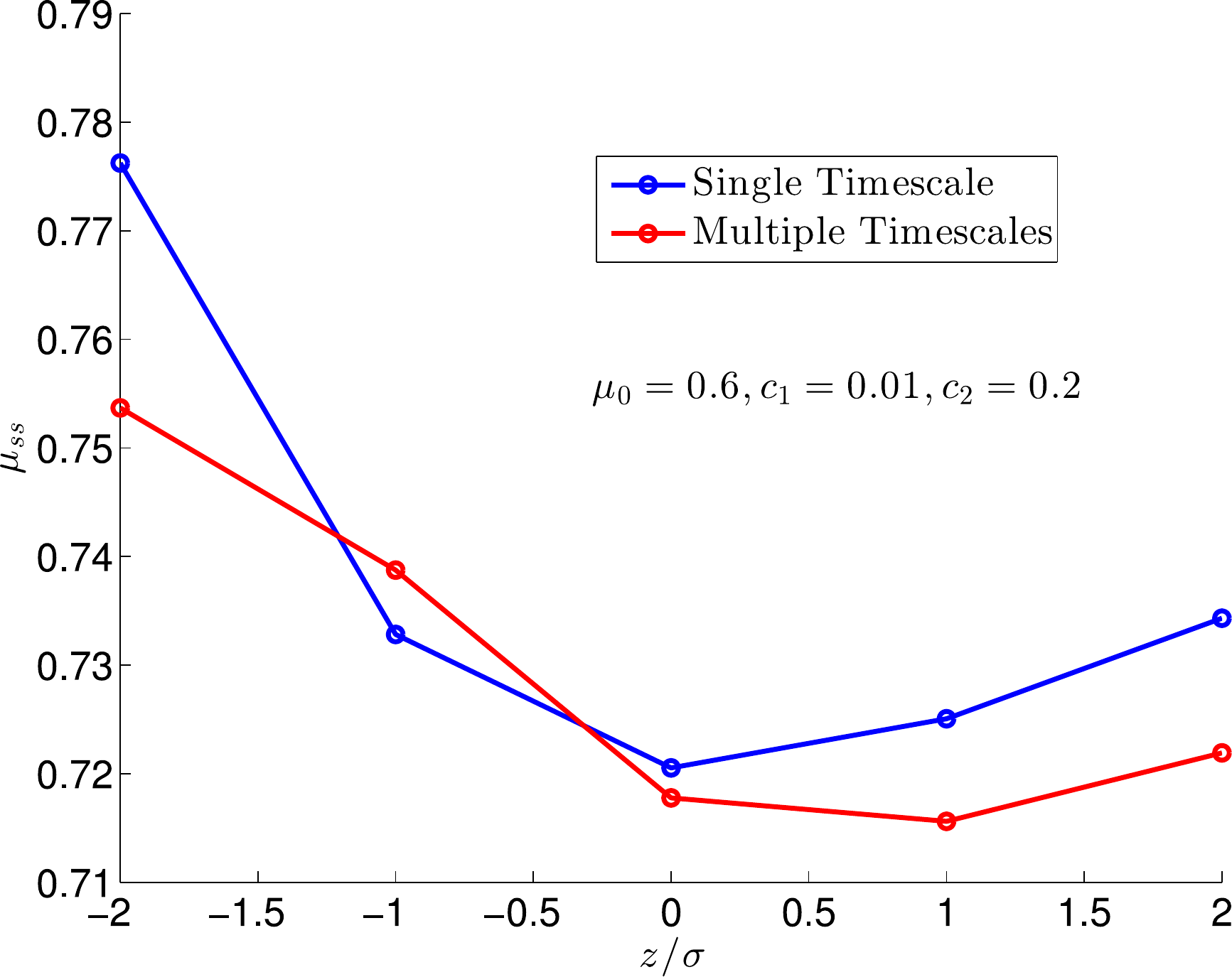}
\subcaption{}
\label{fig:frictionCoefficientSteadyStateVelocityDependenceSingleMultipleTimescales}
\end{subfigure} 
\caption{(a) Evolution of static friction coefficient with hold time for an
ensemble of SLS with two timescales.
(b) The steady-state friction coefficient as a function of the sliding speed
for an ensemble of SLS with two timescales.}
\end{figure}

We conjectured that the saturation of the static friction coefficient 
in $3$-$4$ decades (Figure \ref{fig:frictionEvolutionStaticContactLocalFrictionLaw})
and  the limited velocity range of velocity weakening (Figure 
\ref{fig:frictionCoefficientSteadyStateVelocityDependence}) is primarily 
due to the single timescale in the SLS. To verify this, we perform the 
static contact and velocity jump simulations with an ensemble of SLS 
having multiple timescales. Figure
\ref{fig:frictionEvolutionStaticContactSingleMultipleTimescales} shows the evolution of 
the static friction coefficient in two cases. In the first case (blue), all
sliders in the ensemble have the same relaxation time, $\eta = 1$. In the
other case (red), half the sliders have $\eta = 1$ and the other half have 
$\eta = 10$. The logarithmic growth regime is wider for the
case with two timescales.
Figure \ref{fig:frictionCoefficientSteadyStateVelocityDependenceSingleMultipleTimescales}
shows the steady state friction coefficient as a function of the sliding speed.
The first case (blue) has all sliders with $\eta = 1$ whereas in the second
case (red), half the sliders have $\eta = 1$ while the other half have 
$\eta = 0.1$. We see that the case with two timescales has a broader 
velocity-weakening regime. This confirms our conjecture that the existence of
multiple timescales leads to longer evolution of the static friction 
coefficient and wider velocity-weakening regimes. 

\section{Conclusion}
\label{sec:Conclusion}
We have proposed a model for friction between macroscopic surfaces 
considering asperities at the microscale to be viscoelastic and modeling rough
surfaces as a stochastic process. Our main contribution is a framework to link 
properties of single asperities and surface features to the macroscopic static
and sliding frictional behavior. 
We show that, because of the collective response of contacts, the
behavior at the macroscale can be very different from that at the microscale.

Even in our relatively simple model, rate and state effects appear naturally
as a consequence of the collective asperity response. The model reproduces
the logarithmic strengthening of the static friction with hold time, albeit
within a limited range of time scales compared to some experiments. 
For kinetic friction, the model reproduces many experimental observations.
To capture the direct effect of these experiments,
we need to endow the microscopic  asperity-level friction law with a 
velocity-strengthening dependence. This does not, however, imply that the 
macroscopic response is velocity-strengthening.
The velocity strengthening of individual contacts and the collective behavior
together determine whether the macroscopic response is velocity-strengthening
or weakening. 

The power $n$ in the local friction law plays a crucial role
in the evolution of the friction coefficient. If $n>1$, the static friction
coefficient decreases with hold time and the transient evolution of kinetic 
friction in velocity jump experiments happens in the same sense as the 
instantaneous change. Since we expect the power $n$ to be less than or 
equal to $1$, this explains why the above two features are not 
observed in experiments. 

At the same time, the model appears to be too simple to reproduce all 
experimental observations.  In particular, our model results in velocity 
dependence of the characteristic evolution length scale $D_c$ and parameter
$b$ that quantifies the transient change in friction in the rate and state 
equations, whereas these quantities are largely velocity-independent in 
experimental studies for a range of sliding velocities.  

There are a number of ways to improve the model. In the present model, 
individual asperities are independent, the only interaction between them
being through a mean field (dilatation). Long-range elastic interactions 
through the bulk may, however, play an important role. 
We have also assumed one of the surfaces to be rigid. While this is reasonable when
one of the surfaces is much more deformable than the other, when the two surfaces are 
similar (with respect to deformability), it will be important to incorporate the 
non-rigidity, especially during sliding. 
We have neglected the spatial distribution of asperities and contacts;
this will, however, be important for reproducing realistic frictional behavior,
especially when the long-range interactions are
incorporated. Depending on the material of the sliding surfaces, the model for 
a single asperity can be modified to incorporate effects such as plasticity, 
adhesion etc. These issues remain a topic of current and future work.

\section*{Acknowledgment}
We gratefully acknowledge the support for this study from the National Science
Foundation (grant EAR 1142183) and the Terrestrial Hazards Observations and 
Reporting center (THOR) at Caltech.
We are grateful to Jim Barber, Houman Owhadi, Ronald Fox and Jeff Amelang
for fruitful discussions.

\bibliographystyle{unsrt}
\bibliography{collectiveSlsFriction}

\end{document}